# A Safety and Liveness Theory for Total Reversibility*

Claudio Antares Mezzina[1] and Vasileios Koutavas[2]

[1] IMT Institute for Advanced Studies Lucca
[2] Trinity College Dublin, Dublin, Ireland

**Abstract.** We study the theory of safety and liveness in a reversible calculus where reductions are totally ordered and rollbacks lead the systems to past states. Similar to previous work on communicating transactions, liveness and safety respectively correspond to the should-testing and inverse may-testing preorders. We develop fully abstract models for these preorders in a reversible calculus, which are based only on forward transitions, thus providing a simple proof technique for refinement of such systems. We show that with respect to safety, total reversibility is a conservative extension to CCS. With respect to liveness, however, adding total reversibility to CCS distinguishes more systems. To our knowledge, this work provides the first characterisations of safety and liveness, and the first testing theory for a reversible calculus.

## 1 Introduction

A reversible system is a system that can execute forwards but also backwards, reversing the effects of its computations. Systems that can go back to past states appear in different disciplines, including fault-tolerant systems [8], reverse debugging [9], transactional systems [20], and computational biology [2]. Moreover, there is recent interest in low energy reversible computing based on Landauer's principle [11, 1], which states that only information loss in irreversible computation needs to consume energy.

Despite the wide interest in reversibility [3, 16, 13], its underlying behavioural theory is still not clear. Lanese et al. [13] have observed that, surprisingly, strong barbed congruence [15], the behavioural equivalence most closely related to bisimulation, is a too coarse relation for a reversible calculus (even more in the weak case). On the other hand, if the equivalence distinguishes the direction of reductions (backward or forward), as in the case of back and forth bisimulation [16], the resulting equivalence is too fine as it coincides with history-preserving bisimulation. Thus bisimulation-based relations seem to be unsatisfactory notions of behavioural equivalence in reversibility.

In the presence of non-determinism, many alternatives to bisimulation-based behavioural equivalences exist [19]. The most common ones are may- and must-testing [5] and should-testing [17] preorders. The may-testing preorder is known to correspond to the preservation of *safety* properties. To briefly explain this, consider a safety property $\mathcal{P}$ which expresses that "something bad will not happen" [10]. Safety properties are exactly the properties enforced by monitors [18], which can be thought of as test processes

* This research is supported, in part, by the EU COST Action IC1405. The second author was supported by Science Foundation Ireland grant 13/RC/2094.

running in parallel with the system, reporting property violations on a special channel $\omega$. Thus, if $M \sqsubseteq_{\mathsf{safe}} N$ and $T$ is the monitor of $\mathcal{P}$, if the parallel composition $N \parallel T$ can output on $\omega$ (i.e., one execution of $N$ violates the property monitored by $T$) then $M \parallel T$ can also output on $\omega$. This means that $N \sqsubseteq_{\mathsf{may}} M$ according to may-testing.

Similarly, preservation of liveness properties can be expressed using the must- or the should-testing preorder. In this case tests report on $\omega$ "when something good eventually happens". However there is a subtle difference between the two testing preorders. In must testing, if $M \parallel T$ can diverge without an output on $\omega$, the test fails. The intuition is that $M$ has an execution where the "something good" will not eventually happen. This is a problem in reversibility because a reversible term can infinitely go back and forth between adjacent states, thus it fails every must-test.

On the other hand, $M$ passing a should-test $T$ requires that all future states of $M \parallel T$ have the potential to output on $\omega$ (perhaps after performing some reductions). In a sense this requires that $\omega$ is always *enabled*. If this is the case in an infinite execution, and we assume that infinitely enabled outputs are bound to happen (a fairness assumption), then the success criterion of should testing is justified. Expressing the preservation of liveness properties using should-testing is suitable even for reversible languages. Therefore, in this paper we develop the theory of safety and liveness preservation in terms of the inverse may-testing and should-testing preorders, respectively.

Our model of safety preservation is (inverse) forward trace inclusion, identical to classic CCS. This means that reversibility is a *conservative extension* of CCS with respect to safety. Perhaps not surprisingly, this result relies on the property that any state reachable from an initial system can be also reached with forward moves only. This property can be shown to hold in other reversible calculi (e.g., [3, 4, 12]) where our safety model also applies, but it does not hold in communicating transactions [7].

For liveness preservation, however, it is unclear that models of forward traces are sound. For example the CCS system $M = \nu a.(a.c.\mathbb{0} \parallel a.\mathbb{0} \parallel \overline{a}.\mathbb{0})$ fails the liveness test $T = \overline{c}.\omega$ because $M \parallel T \rightharpoonup \nu a.(a.c.\mathbb{0} \parallel \mathbb{0} \parallel \mathbb{0}) \parallel \overline{c}.\omega$ which is deadlocked and cannot reach $\omega$. On the other hand, system $N = \nu a.(a.c.\mathbb{0} \parallel a(\gamma).\mathtt{rl}\langle\gamma\rangle \parallel \overline{a}.\mathbb{0})$ (where $\gamma$ is a decoration of the $a$-input and $\mathtt{rl}\langle\gamma\rangle$ rolls back this input) passes $T$ because the $a$-reduction that led to deadlock in $M$ can be reversed in $N$.

The question now becomes: does a sound characterisation of the liveness preorder need to consider the full generality of the reversible transition system and explore all states reachable with forward and backward transitions? We answer this question to the negative, providing a simple model for the liveness preorder. This model is based on a novel definition of tree refusals, and *only* requires us to consider the forward transitions of systems and a very limited set of rollback actions. This makes liveness easy to understand and provides a simple proof technique of liveness preservation.

In Sections 2 and 3 we respectively present $\text{CCS}_{\text{roll}}$ and the contextual preorders for safety and liveness. In Section 4 we describe the two crucial semantic properties for the soundness of our forward-transition models of the preorders. In Section 5 we give a compositional semantics in terms of a labelled transition system (LTS) for forward moves and key results on LTS traces necessary for the models of the preorders (Sections 6 and 7). Section 8 contains related work and conclusions. [3]

[3] Technical details of results presented here can be found in an online report [14].

$$\alpha ::= \tau \mid a \qquad \iota ::= k \mid \gamma \qquad a,b,\ldots \in \mathsf{Name} \qquad k,l \in \mathsf{Key} \qquad \gamma,\delta \in \mathsf{KeyVar}$$

$$P,Q ::= \textstyle\sum_{i\in I} \alpha_i(\gamma).P_i \mid P \parallel Q \mid \nu a.P \mid \mathtt{rec}X(\gamma).P \mid X \mid \mathtt{rl}\langle\iota\rangle \qquad (\mathsf{Proc})$$

$$M,N ::= \mathbb{0} \mid \nu a.M \mid M \parallel N \mid k{:}P \mid [\mu;\ k] \qquad (\mathsf{Sys})$$

$$\mu ::= k{:} \textstyle\sum_{i\in I} \alpha_i(\gamma).P_i \mid k{:}\mathtt{rec}X(\gamma).P \qquad (\mathsf{Mem})$$

$$k{:}\nu a.P = \nu a.k{:}P \qquad k{:}(P \parallel Q) = k{:}P \parallel k{:}Q \tag{1}$$

$$\mathbb{0} \parallel A = A \qquad A \parallel B = B \parallel A \qquad (A \parallel B) \parallel C = A \parallel (B \parallel C) \tag{2}$$

$$(\nu a.\mathbb{0}) = \mathbb{0} \qquad (\nu a.A) \parallel B = \nu a.(A \parallel B) \ \ (\text{when } a \sharp B) \qquad \nu a.\nu b.A = \nu b.\nu a.A \tag{3}$$

**Fig. 1.** $\text{CCS}_{\text{roll}}$ Syntax and Structural Equivalence Axioms.

## 2 The Language $\text{CCS}_{\text{roll}}$

The language $\text{CCS}_{\text{roll}}$ extends CCS with a form of controlled reversibility, where reductions are totally ordered and systems can be programmed to return to any previous state. As in CCS, synchronisation between processes occurs over channel names (Name) according to a total, irreflexive bijection ($\overline{\cdot}$) over Name. Unlike CCS, unique *keys* (Key) are used to identify and roll back synchronisation and internal ($\tau$) reductions.

The $\text{CCS}_{\text{roll}}$ syntax is shown in fig. 1 and is organised in two levels: *processes* and *systems*. Processes (Proc) include standard CCS processes for infinite choice operator $\sum_{i\in I} \alpha_i(\gamma).P_i$, where $I$ is an indexing set[4]; parallel composition $P \parallel Q$; name restriction $\nu a.P$; recursive process definition, process variable and a rollback primitive $\mathtt{rl}\langle\iota\rangle$. A key variable $\gamma$ decorates both prefix and recursion. When a prefix is reduced (or a recursion in unfolded), $\gamma$ is replaced with a fresh key $k$ in the continuation. From this point, the process can roll back to the state before this reduction by executing $\mathtt{rl}\langle k\rangle$.

Systems (Sys) are build up from *named processes* $k{:}P$ and *memories* $[\mu;\ k]$. In $k{:}P$ the key $k$ identifies the reduction that produced $P$, which may be shared with other processes produced by the same reduction. A memory $[\mu;\ k]$ records that the named process $\mu$ was involved in the $k$-reduction. If a past $k$-reduction was a synchronisation, the current system will contain two memories $[\mu_1;\ k]$ and $[\mu_2;\ k]$ recording the named processes that synchronised. Otherwise the $k$-reduction was internal (i.e. either a $\tau$ prefix or a recursive process unfold) and the system will contain only one memory of the form $[\mu;\ k]$. We make use of two structural equivalences over $A, B, C \in \mathsf{Sys} \cup \mathsf{Proc}$.

**Definition 1 (Structural Equivalences).** Limited structural equivalence ($\mathrel{\mathring{=}}$) *for systems is defined to be the least equivalence satisfying the axioms (1) of fig. 1 and closed under parallel* ($- \parallel -$) *and name restriction* ($\nu a.-$). Structural equivalence ($\equiv$) *is obtained by also requiring the axioms (2) and (3) of fig. 1.*

The reduction semantics of $\text{CCS}_{\text{roll}}$ is defined by transitions between configurations of the form $D \vdash M$, where $D$ is a *dependency history*, recording the order of reductions and derived by the following grammar:

$$D ::= \varepsilon \mid k \prec D \qquad (\text{if } k \sharp D)$$

[4] By convention when $I = \emptyset$ we have $\sum_{i\in I} \alpha_i(\gamma).P_i = \mathbb{0}$.

$$\text{R}\alpha\ \frac{j \in I \qquad k \sharp (P_i)_{i \in I}}{\sum_{i \in I} \alpha_i(\gamma).P_i \xmapsto{\alpha_j(k)} P_j\{^k/_\gamma\}} \qquad \text{RREC}\ \frac{k \sharp P}{\mathtt{rec}X(\gamma).P \xmapsto{\tau(k)} P'\{^{\mathtt{rec}X(\gamma).P}/_X\}\{^k/_\gamma\}}$$

$$\text{RSYNC}\ \frac{P \xmapsto{a(k)} P' \qquad Q \xmapsto{\overline{a}(k)} Q' \qquad k \sharp (k_1, k_2, P, Q)}{k_1{:}P \parallel k_2{:}Q \stackrel{(k)}{\rightharpoonup} k{:}P' \parallel k{:}Q' \parallel [k_1{:}P;\ k] \parallel [k_2{:}Q;\ k]}$$

$$\text{R}\tau\ \frac{P \xmapsto{\tau(k)} P' \qquad k \sharp k_0}{k_0{:}P \stackrel{(k)}{\rightharpoonup} k{:}P' \parallel [k_0{:}P;\ k]} \qquad \text{RPAR}\ \frac{M \stackrel{(k)}{\rightharpoonup} M' \qquad k \sharp N}{M \parallel N \stackrel{(k)}{\rightharpoonup} M' \parallel N}$$

$$\text{R}\nu\ \frac{M \stackrel{(k)}{\rightharpoonup} M'}{\nu a.M \stackrel{(k)}{\rightharpoonup} \nu a.M'} \qquad \text{REQV}\ \frac{M \equiv N \quad N \stackrel{(k)}{\rightharpoonup} N' \quad N' \equiv M'}{M \stackrel{(k)}{\rightharpoonup} M'}$$

$$\text{RSYS}\ \frac{M \stackrel{(k)}{\rightharpoonup} M' \quad k \sharp D}{D \vdash M \stackrel{(k)}{\rightharpoonup} k \prec D \vdash M'} \qquad \text{FW}\ \frac{D \vdash M \stackrel{(k)}{\rightharpoonup} D' \vdash M'}{D \vdash M \rightharpoonup D' \vdash M'}$$

**Fig. 2.** Reduction Semantics: Forward Rules.

The side-condition guarantees that each $k$ in a dependency history is recorded at most once. We write $D \prec D'$ for the concatenation of $D$ and $D'$, and $\widetilde{k} \prec D$ for $k_1 \prec \ldots k_n \prec D$. We also write $k \prec_D l$ when $D = D_1 \prec k \prec l \prec D_2$. When the dependency history $D$ is clear from the context we will write $k \prec l$ instead of $k \prec_D l$.

Before a configuration starts running, it contains no memories and no keys corresponding to past reductions. We call these configurations *initial*. To abide to the syntax for systems we consider one special key $\varepsilon$ which is used to annotate processes in initial systems of the form $M \mathrel{\mathring{=}} \varepsilon{:}P$. A configuration $D \vdash M$ is initial if $M \mathrel{\mathring{=}} \varepsilon{:}P$ and $D = \varepsilon$. In the following we use standard CCS abbreviations, such as $\widetilde{o}$ for a sequence $o_1, \ldots, o_n$, where $n$ is implicit. We write $\mathsf{keys}(o)$ for the set of keys in $o$ except for $\varepsilon$, and $o_1 \sharp o_2$ when $\mathsf{keys}(o_1) \cap \mathsf{keys}(o_2) = \emptyset$. We omit variables $\gamma$ when not used.

The reduction relation of $\text{CCS}_{\text{roll}}$ $\rightarrow$ (over configurations) is defined as the union of the *forward* reduction relation $\rightharpoonup$ and the *backward*, or rollback, relation $\hookrightarrow$. Relations $\rightharpoonup$ and $\hookrightarrow$ are given in figs. 2 and 3.

A forward reduction takes the form $D \vdash M \rightharpoonup D' \vdash M'$, and it is derived from rules FW and RSYS of fig. 2 only when $D' = k \prec D$ and $M \stackrel{(k)}{\rightharpoonup} M'$. The key $k$ is fresh ($k \sharp D, M$) and uniquely identifies the reduction. The rules deriving $M \stackrel{(k)}{\rightharpoonup} M'$ are adapted from CCS. Note that rule (RSYNC) generates two memories, each one containing the named process contributing to the synchronisation and the fresh key $k$. Rule (R$\tau$) generates one memory containing the named process performing the internal action along with the key of the reduction. When a prefix is reduced (via rule R$\alpha$ or RREC) the bound variable $\gamma$ is substituted with the new key in the continuation of the process, replacing free occurrences of $\gamma$ in terms $\mathtt{rl}\langle\gamma\rangle$, allowing the roll-back of the reduction.

A backward reduction $D \vdash M \hookrightarrow D' \vdash M'$ is derived from rule BW of fig. 3 only when a rollback of $k$ is *active* in $M$ (i.e. $M \equiv l{:}\mathtt{rl}\langle k\rangle \parallel N$). In $\text{CCS}_{\text{roll}}$ actions are

$$\textsc{RLmem1}\quad \frac{}{[\mu;\,k] \xhookrightarrow{\langle k\rangle} \mu} \qquad \textsc{RLmem2}\quad \frac{k \sharp k', \mu}{[\mu;\,k'] \xhookrightarrow{\langle k\rangle} [\mu;\,k']} \qquad \textsc{RLproc1}\quad \frac{}{k{:}P \xhookrightarrow{\langle k\rangle} \mathbb{0}} \qquad \textsc{RLproc2}\quad \frac{k \sharp k', P}{k'{:}P \xhookrightarrow{\langle k\rangle} k'{:}P}$$

$$\textsc{RLnil}\quad \frac{}{\mathbb{0} \xhookrightarrow{\langle k\rangle} \mathbb{0}} \qquad \textsc{RL}\nu\quad \frac{M \xhookrightarrow{\langle k\rangle} M'}{\nu a.M \xhookrightarrow{\langle k\rangle} \nu a.M'} \qquad \textsc{RLpar}\quad \frac{M \xhookrightarrow{\langle k\rangle} M' \qquad N \xhookrightarrow{\langle k\rangle} N'}{M \parallel N \xhookrightarrow{\langle k\rangle} N' \parallel M'}$$

$$\textsc{RLsys}\quad \frac{M \xhookrightarrow{\langle k_1\rangle} \dots \xhookrightarrow{\langle k_n\rangle} M'}{(k_1 \prec \ldots \prec k_n \prec D) \vdash M \xhookrightarrow{\langle k_n\rangle} D \vdash M'} \qquad \textsc{Bw}\quad \frac{M \equiv l{:}\mathtt{rl}\langle k\rangle \parallel N \qquad D \vdash M \xhookrightarrow{\langle k\rangle} D' \vdash M'}{D \vdash M \hookrightarrow D' \vdash M'}$$

**Fig. 3.** Reduction Semantics: Backward Rules.

totally ordered, and their identifying keys are stored in the dependency history $D$. Reverting a $k$-action reverts all actions that came after it (RLsys). Each reduction of the form $M \xhookrightarrow{\langle k\rangle} N$ is derived from the rules in the first part of fig. 3, which broadcast the $k$-rollback throughout $M$. Processes and memories in $M$ that do not contain $k$ are left unaffected by this transition; $k$-processes disappear and $k$-memories reinstate their contents. We write $\xrightarrow{(\widetilde{k})}$ and $\xhookrightarrow{\langle\widetilde{k}\rangle}$ for $\xrightarrow{(k_1)} \dots \xrightarrow{(k_n)}$ and $\xhookrightarrow{\langle k_1\rangle} \dots \xhookrightarrow{\langle k_n\rangle}$, respectively.

In $\mathrm{CCS_{roll}}$, rollbacks are deterministic. Moreover, any forward $k$-reduction can be rolled back, and when this happens the configuration returns to the state before the $k$-reduction, up to structural equivalence.

**Lemma 1 (Deterministic Rollback).** *Let $D \vdash M \xhookrightarrow{\langle k\rangle} D' \vdash M'$ and $D \vdash M \xhookrightarrow{\langle k\rangle} D'' \vdash M''$. Then $D' = D''$ and $M' = M''$.*

**Lemma 2 (Rollback a Forward Reduction).** *If $D \vdash M \xrightarrow{(k)}\xrightarrow{(\widetilde{l})} D' \vdash M'$ then there exists $M''$ such that $D' \vdash M' \xhookrightarrow{\langle k\rangle} D \vdash M''$ and $M \equiv M''$.*

To establish more properties for configurations we require them to be *well-formed*.

**Definition 2 (Well-Formed Config.).** *$D \vdash M$ is well-formed, written $\mathsf{wf}(D \vdash M)$, if*

1. Key Compatibility: $\mathsf{keys}(M) \subseteq \mathsf{keys}(D)$,
2. Rollback Loop: *if $D = k \prec D'$ then $M \xhookrightarrow{\langle k\rangle} M' \xrightarrow{(k)} M$ and $\mathsf{wf}(D' \vdash M')$, for some $M'$.*

This definition guarantees that in a configuration $D \vdash M$, keys in $M$ were produced by a past reduction recorded in $D$, and any such reduction can be rolled back and repeated obtaining the same system. These properties are sufficient to describe well-behaved systems, simplifying the definitions of previous work [12].

Well-formedness is preserved by structural equivalence and reductions. This, together with the fact that initial configurations are trivially well-formed, allows us to implicitly assume configuration to be well-formed in the following sections.

| | |
|---|---|
| $M_1 = \varepsilon{:}a.(b.\mathbb{0} + c.\mathbb{0})$ | $M_2 = \varepsilon{:}(a.b.\mathbb{0} + a.c.\mathbb{0})$ |
| $M_3 = \varepsilon{:}a.(b.c.\mathbb{0} + b.d.\mathbb{0})$ | $M_4 = \varepsilon{:}(a.b.c.\mathbb{0} + a.b.d.\mathbb{0})$ |
| $M_5 = \varepsilon{:}a.(b(\gamma_b).\mathtt{rl}\langle\gamma_b\rangle + c.\mathbb{0})$ | $M_6 = \varepsilon{:}a(\gamma_a).(b.\mathtt{rl}\langle\gamma_a\rangle + c.\mathbb{0})$ |

**Fig. 4.** Examples.

# 3 Safety and Liveness Preorders

In this section we give the definitions of the safety and liveness preorders, and examples of their use. As we discussed in the introduction, the safety preorder corresponds to the inverse may-testing preorder [5] and the liveness preorder corresponds to the should-testing preorder [17]. Here we use tests $T$ derived from the grammar of processes, with the addition of a special name $\omega$ used by the test to report an outcome.

**Definition 3 (Basic Observable (barb)).** *$D \vdash M$ has a* strong barb*, written $D \vdash M{\downarrow_\omega}$, if $M \equiv N \parallel k : \omega$ and a* (weak) barb *written $D \vdash M{\Downarrow_\omega}$, if $D \vdash M \twoheadrightarrow^* D' \vdash M'{\downarrow_\omega}$.*

We are interested in testing initial configurations; the composition of a configuration $\varepsilon \vdash M$ with a test $T$ is $\varepsilon \vdash M \parallel \varepsilon{:}T$. We start with safety. A safety test $T$ can be thought of as a monitor enforcing a safety policy. When $T$ reports an error on $\omega$ then the enforced safety policy has been violated by the system. Thus, a system $M$ passes a safety test $T$ when their parallel composition cannot report a violation on $\omega$. This, in negative form, means that $M$ fails $T$ if $\varepsilon \vdash M \parallel \varepsilon{:}T{\Downarrow_\omega}$. System $M$ is potentially "less safe" than $N$ when $M$ has at least the violations of $N$.

**Definition 4 (Safety Preorder).** *For two initial systems $M$ and $N$ we write $M \sqsubseteq_{\mathsf{safe}} N$ when for all tests $T$, $\varepsilon \vdash N \parallel \varepsilon{:}T{\Downarrow_\omega}$ implies $\varepsilon \vdash M \parallel \varepsilon{:}T{\Downarrow_\omega}$.*

On the other hand, we can consider *liveness tests* that report on $\omega$ when "something good" happens. A system $M$ passes a liveness test $T$ when their parallel composition has no way of failing the test. According to should-testing [17], a system passes a test if at any reachable state, success is reachable; $M$ is potentially "less live" than $N$ if $N$ passes every liveness test that $M$ passes.

**Definition 5 (Passing a Liveness Test & Liveness Preorder).** *An initial system $M$ passes the liveness test $T$, written $M$ $\mathsf{shd}$ $T$, when $\varepsilon \vdash M \parallel \varepsilon{:}T \twoheadrightarrow^* D \vdash N$ implies $D \vdash N{\Downarrow_\omega}$, $\forall D, N$. For two initial systems $M$ and $N$ we write $M \sqsubseteq_{\mathsf{live}} N$ when for all liveness tests $T$, $M$ $\mathsf{shd}$ $T$ implies $N$ $\mathsf{shd}$ $T$.*

In the example systems in fig. 4, $M_1$ and $M_2$ are CCS processes. In CCS these are safe-equivalent because they have the same traces. With respect to the liveness preorder, also in CCS, $M_2 \sqsubseteq_{\mathsf{live}} M_1$ but $M_1 \not\sqsubseteq_{\mathsf{live}} M_2$ because $M_1$ passes $T = \omega + \overline{a}.\overline{b}.\omega$ but $M_2$ does not: after a communication on $a$, $M_2 \parallel T$ can become $c.\mathbb{0} \parallel \overline{b}.\omega$, which cannot reach $\omega$. The same is true in $\text{CCS}_{\text{roll}}$. The following conservative extension theorem holds for safety in $\text{CCS}_{\text{roll}}$– its proof relies in the models for safety in the two languages.

**Theorem 1.** *Let $P$ and $Q$ be $CCS$ processes with $P \sqsubseteq^{\mathsf{CCS}}_{\mathsf{safe}} Q$; then $\varepsilon{:}P \sqsubseteq^{CCS_{roll}}_{\mathsf{safe}} \varepsilon{:}Q$.*

Terms $M_3$ and $M_4$ of fig. 4 show why CCS$_{\text{roll}}$ is *not* a conservative extension of CCS with respect to liveness. In CCS $M_3$ is safe-equivalent to $M_4$ [17, Ex. 34], however these terms can be distinguished in CCS$_{\text{roll}}$ by test $T = \omega + \overline{a}.\tau(\gamma).(\mathtt{rl}\langle\gamma\rangle \parallel \overline{b}.\overline{d}.\omega)$. System $M_3$ passes this test because if, while communicating with $T$, it chooses the "wrong" $b$-branch $b.d.\mathbb{0}$, the rollback in the test reverses this choice and $\omega$ is reachable. On the other hand, $M_4$ can communicate with $T$ using branch $a.b.d.\mathbb{0}$ where $\omega$ will not be reachable; the rollback in the test cannot reverse this choice. For a similar reason, $M_5$ and $M_6$ are safe- but not live-equivalent. System $M_6$ passes the liveness test $T = \omega + a.b$ but $M_5$ does not; the rollback of the $a$-communication in $M_6$ makes $\omega$ to always be reachable.

## 4 Important Properties of Forward Reductions

Our safety and liveness models rely on two semantic properties. The first is that, starting from any system, any state reachable with arbitrary reductions can also be reached with only forward reductions after at most one backward move. Thus, all states of an initial system, which has no backward moves, can be reached with only forward reductions.

The consequence of this is that forward traces are sufficient to characterise safety. This property also holds in any calculus whose reversibility machinery is causally consistent, thus our forward-only model of safety applies to other reversible systems.

**Lemma 3.** *Let $D \vdash M \rightarrow^* D' \vdash M'$. Then one of the following holds:*

1. $D \vdash M \rightharpoonup^* D' \vdash M'' \equiv M'$*, for some* $M''$*, or*
2. $D \vdash M \xhookrightarrow{\langle l \rangle} D_0 \vdash M_0 \rightharpoonup^* D' \vdash M'' \equiv M'$*, and* $D \vdash M \rightarrow^* D_0 \vdash M_0' \equiv M_0$*, for some* $l \in D$*,* $D_0$*,* $M''$*,* $M_0$*,* $M_0'$*.*

**Property 1** *Let $\varepsilon \vdash M \rightarrow^* D' \vdash M'$. Then $\varepsilon \vdash M \rightharpoonup^* D' \vdash M'' \equiv M'$, for some $M''$.*

This property is also necessary for the characterisation of liveness. However we also need to establish a result for tree failures [17]: if an initial configuration can reach a state $D \vdash N$ from which it fails to reach an $\omega$-action, then the same original configuration should be able to reach a failure state $D_1 \vdash N_1$ where all reachable states can be reached with forward reductions. This allows us to use a forward LTS to encode liveness.

**Property 2** *Let $\varepsilon \vdash M \rightharpoonup^* D \vdash N \not\Downarrow_\omega$ and $\mathsf{keys}(M) = \emptyset$; there exist $D_1$ and $N_1$:*

1. $\varepsilon \vdash M \rightharpoonup^* D_1 \vdash N_1 \not\Downarrow_\omega$
2. *if* $D_1 \vdash N_1 \rightarrow^* D' \vdash N'$ *then there exists* $N''$ *such that* $D_1 \vdash N_1 \rightharpoonup^* D' \vdash N'' \equiv N'$

In CCS$_{\text{roll}}$, where reductions are temporally dependent, we can show that once we reach the state $D \vdash N$ we can explore all past states reachable with rollbacks from $D \vdash N$ and pick the oldest one. Because of the total temporal ordering of reductions we know that there is always a single oldest state (up to structural equivalence $\equiv$) $D_1 \vdash N_1$. Since no older state is reachable, all states reachable from $D_1 \vdash N_1$ can be reached with only forward reductions.

$$\textsc{T}\alpha \quad \frac{P \xmapsto{\alpha(k)} P' \qquad k \sharp D}{D \models l{:}P \xrightarrow{\alpha(k)} (k \prec D) \models k{:}P' \parallel [l{:}P;\, k]} \qquad \textsc{Tsync} \quad \frac{D \models M \xrightarrow{a(k)} D' \models M' \qquad D \models N \xrightarrow{\overline{a}(k)} D' \models N'}{D \models M \parallel N \xrightarrow{\tau(k)} D' \models M' \parallel N'}$$

**Fig. 5.** LTS Transitions Based on Forward Moves.

## 5 Compositional Semantics

Our characterisation of safety and liveness in $\text{CCS}_{\text{roll}}$ is based on a compositional Labelled Transition System (LTS) of forward transitions between compositional configurations of the form $D \models M$, greatly simplifying reasoning. We let $\mathcal{C}$ range over compositional configurations. The LTS transition relation $\xrightarrow{\alpha(k)}$ is defined as the smallest relation derived from fig. 5, closed under $\stackrel{\circ}{=}$, parallel and restriction, provided that $k$ in the label is fresh.

These transitions, besides internal reductions, can describe the interaction of a *part* of a system with its environment, which we call the observer. We assume the adaptation of the definition of well-formed configurations to compositional configuration and work with well-formed compositional configurations. As expected, forward reductions correspond to $\tau$-transitions in the LTS.

**Theorem 2.** $D \vdash M \xrightarrow{(k)} D' \vdash M'$ *iff* $D \models M \xrightarrow{\tau(k)} D' \models M''$ *with* $M'' \equiv M'$.

Our theory is based on *canonical traces*; $t$ is canonical if each key in $t$ appears at most once in $t$. A trace is a dependency history transformer and can be typed as such.

**Definition 6 (Trace Typing).** *We write* $(D \vdash t \triangleright D')$ *for the predicate defined by the following rules:*

$$(D \vdash \epsilon \triangleright D) \qquad\qquad (D \vdash \alpha(k), t \triangleright D') \;\; \textit{if} \;\; (k \prec D \vdash t \triangleright D')$$

We will treat $(D \vdash t \triangleright D')$ as a typed trace; this formalism helps us to synchronise dependency histories with traces. Canonical traces are typable, provided they use new keys, and any typed trace is canonical.

Traces encode both the observable and internal ($\tau$) actions of a system. Systems related by the safety and liveness preorders may have traces that differ in their internal actions. We write $\mathsf{obs}(t)$ to denote the sub-trace of $t$ containing only observable actions.

$$\mathsf{obs}(\epsilon) = \epsilon \qquad \mathsf{obs}(\tau(k), t) = \mathsf{obs}(t) \qquad \mathsf{obs}(a(k), t) = a(k), \mathsf{obs}(t)$$

We say that a trace $t$ is *observable* when $\mathsf{obs}(t) = t$. We write $\overline{t}$ to denote the same-length trace derived from $t$ by applying $(\overline{\cdot})$ to all non-$\tau$ actions. If $t_1 = \overline{t}_2$ then we call $t_1$ and $t_2$ *complementary*. A single LTS transition of two parallel systems can be decomposed to either a transition of one of the systems, or a synchronisation between them. This leads to a general decomposition of the trace of two parallel systems.

$$\begin{array}{lll}
\varepsilon \parallel \varepsilon & \twoheadrightarrow \varepsilon & (\mathrm{Z}\varepsilon) \\
D_1 \parallel (k \prec D_2) & \twoheadrightarrow k \prec D \text{ if } D_1 \parallel D_2 \twoheadrightarrow D \text{ and } k \sharp D_1, D_2 & (\mathrm{ZR}) \\
(k \prec D_1) \parallel (k \prec D_2) & \twoheadrightarrow k \prec D \text{ if } D_1 \parallel D_2 \twoheadrightarrow D \text{ and } k \sharp D_1, D_2 & (\mathrm{Z\scriptsize{SYNC}})
\end{array}$$

$$\mathrm{ZT}\epsilon\ \frac{D_1 \parallel D_2 \twoheadrightarrow D}{(D_1 \vdash \epsilon \rhd D_1) \parallel (D_2 \vdash \epsilon \rhd D_2) \twoheadrightarrow (D \vdash \epsilon \rhd D)}$$

$$\mathrm{ZTR}\ \frac{(D_1 \vdash t_1 \rhd D_1') \parallel (k \prec D_2 \vdash t_2 \rhd D_2') \twoheadrightarrow (k \prec D \vdash t \rhd D')}{(D_1 \vdash t_1 \rhd D_1') \parallel (D_2 \vdash \alpha(k), t_2 \rhd D_2') \twoheadrightarrow (D \vdash \alpha(k), t \rhd D')}$$

$$\mathrm{ZTS\scriptsize{YNC}}\ \frac{(k \prec D_1 \vdash t_1 \rhd D_1') \parallel (k \prec D_2 \vdash t_2 \rhd D_2') \twoheadrightarrow (k \prec D \vdash t \rhd D')}{(D_1 \vdash a(k), t_1 \rhd D_1') \parallel (D_2 \vdash \overline{a}(k), t_2 \rhd D_2') \twoheadrightarrow (D \vdash \tau(k), t \rhd D')}$$

**Fig. 6.** Definition of Zipping Traces and Histories (Symmetric rules of ZTR and ZR are omitted).

**Proposition 1 (Unzipping).** *Let* $D \models M \parallel N \xrightarrow{t} D' \models R'$ *and* $\mathsf{obs}(t) = \epsilon$ *and* $D_1 \parallel D_2 \twoheadrightarrow D$*. There exist* $M'$, $N'$, $D_1'$, $D_2'$, $t_1$, $t_2$ *such that*

$$D \models M \xrightarrow{t_1} D_1' \models M' \qquad (D_1 \vdash t_1 \rhd D_1') \parallel (D_2 \vdash t_2 \rhd D_2') \twoheadrightarrow (D \vdash t \rhd D')$$

$$D \models N \xrightarrow{t_2} D_2' \models N' \qquad R' \mathrel{\mathring{=}} M' \parallel N' \qquad \mathsf{obs}(t_1) = \overline{\mathsf{obs}(t_2)}$$

Conversely, if two systems can perform typed traces that can be zipped into a single zipped trace, then the parallel composition of these systems can perform this trace.

**Proposition 2 (Zipping).** *Let* $D_1 \models M \xrightarrow{t_1} D_1' \models M'$ *and* $D_2 \models N \xrightarrow{t_2} D_2' \models N'$ *and* $(D_1 \vdash t_1 \rhd D_1') \parallel (D_2 \vdash t_2 \rhd D_2') \twoheadrightarrow (D \vdash t \rhd D')$*. Then* $D \models M \parallel N \xrightarrow{t} D' \models R' \mathrel{\mathring{=}} M' \parallel N'$*.*

## 6 Model of the Safety Preorder

We now show that the safety preorder coincides with inverse observable trace inclusion.

**Definition 7 (Trace Set).** *We write* $\mathsf{Tr}(\mathcal{C})$ *for the largest set of* observable traces *such that* $t \in \mathsf{Tr}(\mathcal{C})$ *when there exists* $t'$ *and* $\mathcal{C}'$ *such that* $\mathsf{obs}(t') = t$ *and* $\mathcal{C} \xrightarrow{t'} \mathcal{C}'$*.*

Observable traces correspond to the class of safety tests defined by the rules:

$$\mathsf{Test}^{\mathsf{s}}(\epsilon) = \omega \qquad \mathsf{Test}^{\mathsf{s}}(a(k), t) = \overline{a}.\mathsf{Test}^{\mathsf{s}}(t)$$

**Lemma 4.** *Let* $t$ *be an observable trace; then*

1. *there exist* $D$, $T$ *such that* $\varepsilon \vdash \mathsf{Test}^{\mathsf{s}}(t) \xrightarrow{\overline{t}} D \vdash T{\downarrow_\omega}$*;*
2. *if* $\varepsilon \vdash \mathsf{Test}^{\mathsf{s}}(t) \xrightarrow{t'} D \vdash T{\downarrow_\omega}$ *then there exists a permutation* $p$ *such that* $\overline{t} = pt'$*.*

Our characterisation of the safety preorder is the inverse of the following *trace preorder*.

**Definition 8 (Trace Preorder).** *For initial systems* $M \sqsubset\!\!\sim_{\mathsf{tr}} N$ *if* $\mathsf{Tr}(\epsilon \models M) \subseteq \mathsf{Tr}(\epsilon \models N)$*.*

**Theorem 3 (Soundness and Completeness).** $M \sqsubset\!\!\sim_{\mathsf{tr}} N$ *iff* $N \sqsubset\!\!\sim_{\mathsf{safe}} M$*.*

It is now easy to check that systems $M_1$, $M_2$, $M_5$ and $M_6$ from fig. 4 are pairwise safe-equivalent because they have the same observable traces, and so are $M_3$ and $M_4$.

## 7 Model of the Liveness Preorder

Our model of the liveness preorder is based on forward traces but includes the following basic observable for rollback actions, determined entirely by the structure of terms.

**Definition 9 (Rollback Barb).** *If $\Delta$ is a set of keys, we write $D \models M{\downarrow_{\mathtt{rl}\langle \Delta \rangle}}$ when $\exists k \in \Delta$ such that $M \equiv l{:}\mathtt{rl}\langle k \rangle \parallel N$, for some $l$, $N$. We let $D \models M{\downarrow_{\mathtt{rl}\langle D' \rangle}}$ mean $D \models M{\downarrow_{\mathtt{rl}\langle \mathsf{keys}(D') \rangle}}$.*

Based on this basic observable for rollbacks, we define the set of observable traces of a configuration $\mathcal{C}$ which lead to a rollback of an action before $\mathcal{C}$.

**Definition 10 (Rollback Traces).** $\mathsf{Roll}(D \models M)$ *is the set of observable traces for which $t \in \mathsf{Roll}(D \models M)$ iff there exists $t'$ such that $\mathsf{obs}(t') = t$ and $D \models M \xrightarrow{(t')} D' \models M'{\downarrow_{\mathtt{rl}\langle D \rangle}}$.*

**Lemma 5.** $\mathsf{Roll}(D \models M) \subseteq \mathsf{Tr}(D \models M)$.

The main structure of our liveness model is a *tree refusal* [17] adapted for reversibility.

**Definition 11 (Tree Refusal).** *A tree refusal is a tuple $(t; V; W)$, where:*

1. *$t$ is an observable trace, and $V$ and $W$ are sets of observable traces,*
2. *$\epsilon \in V$ and $V$ is prefix-closed,*
3. *$\epsilon \notin W$ and $W \subseteq V$.*
4. *$V$ and $W$ are closed under permutation of keys.*

A tree refusal $(t; V; W)$ encodes how an initial system $M$ can fail a liveness test $T$:

- $M$ communicates with $T$ according to the actions in $t$ and together reach the state $D \vdash N$. From this state, an $\omega$ output is not reachable and the liveness test fails.
- At this state the test is offering to communicate on the traces in $W$ and then output on $\omega$. Thus, the system cannot perform the traces in $W$. Since any system can perform the empty trace, $\epsilon \notin W$.
- If the state $D \vdash N$ is rolled back the test will reach an $\omega$ output. Moreover, at this state, all the traces that the test is offering to communicate are in $V$, including those in $W$; note that every test offers the empty trace ($\epsilon \in V$). Thus the system should not be able to roll back state $D \vdash N$ while communicating with the test over $V$. If the test can roll back $D \vdash N$ over a trace $t_0$, we add $t_0 \in W$.

Tree refusals correspond to the following class of *characteristic liveness tests*.

**Definition 12 (Characteristic Liveness Tests).**

$$\mathsf{Test}^{\mathsf{l}}(\epsilon; V; W) \stackrel{\mathit{def}}{=} \omega + \tau.\,\mathsf{Test}^{\mathsf{l}}(V; W) \qquad \mathsf{Test}^{\mathsf{l}}(a(k), t; V; W) \stackrel{\mathit{def}}{=} \omega + \overline{a}.\mathsf{Test}^{\mathsf{l}}(t; V; W)$$

$$\mathsf{Test}^{\mathsf{l}}(V; W) \stackrel{\mathit{def}}{=} \tau(\gamma).\left(\left(\left(\sum_{t_1 \in (V \setminus W)} \mathsf{Test}^{\mathsf{l}}(t_1; \mathbb{0})\right) + \left(\sum_{t_2 \in W} \mathsf{Test}^{\mathsf{l}}(t_2; \omega)\right)\right) \parallel \mathtt{rl}\langle \gamma \rangle\right)$$

$$\mathsf{Test}^{\mathsf{l}}(\epsilon; P) \stackrel{\mathit{def}}{=} P \qquad \mathsf{Test}^{\mathsf{l}}(a(k).t; P) \stackrel{\mathit{def}}{=} \overline{a}.\mathsf{Test}^{\mathsf{l}}(t; P)$$

An initial system fails $\mathsf{Test}^{\mathsf{l}}(t;V;W)$ only by communicating with the test along the trace $t$, thus reducing the test to $\mathsf{Test}^{\mathsf{l}}(V;W)$. From this state, an $\omega$ is reachable only if the system communicates with the test along a trace $t_2 \in W$, or if the system rolls back to a previous state along any of the traces in $V$ which contains $W$. The $\mathtt{rl}\langle\gamma\rangle$ is used to non-deterministically bring the test to state $\mathsf{Test}^{\mathsf{l}}(V;W)$, thus avoiding the deadlock of system and test while communicating along a strict prefix of a trace in $V$.

**Definition 13 (Refusal Set).** *Let $M$ be an initial system; $\mathsf{Ref}(M)$ is the largest set of tree refusals with the property that if $(t;V;W) \in \mathsf{Ref}(M)$, there exist $t'$ and $\mathcal{C}$ s.t.:*

$$\mathsf{obs}(t') = t \qquad \textit{and} \qquad \varepsilon \models M \xrightarrow{(t')} \mathcal{C} \qquad \textit{and} \qquad V \cap \mathsf{Roll}(\mathcal{C}) = W \cap \mathsf{Tr}(\mathcal{C}) = \emptyset$$

**Lemma 6 (Refusals to Traces).** *If $(t;V;W) \in \mathsf{Ref}(M)$ then $t \in \mathsf{Tr}(M)$.*

**Definition 14 (Refusal Preorder).** *For initial systems, $M \sqsubseteq_{\mathsf{ref}} N$ if $\mathsf{Ref}(M) \subseteq \mathsf{Ref}(N)$.*

**Theorem 4 (Soundness and Completeness).** *$M \sqsubseteq_{\mathsf{live}} N$ iff $N \sqsubseteq_{\mathsf{ref}} M$.*

Let us revisit fig. 4: $M_1 \not\sqsubseteq_{\mathsf{live}} M_2$, $M_3 \not\sqsubseteq_{\mathsf{live}} M_4$ and $M_5 \not\sqsubseteq_{\mathsf{live}} M_6$ because

$$\begin{aligned}
(a;\{\epsilon,\ b\};\{b\}) &\in \mathsf{Ref}(M_2) & (a;\{\epsilon,\ b\};\{b\}) &\notin \mathsf{Ref}(M_1)\\
(a;\{\epsilon,\ b,\ b.c\};\{b.c\}) &\in \mathsf{Ref}(M_4) & (a;\{\epsilon,\ b,\ b.c\};\{b.c\}) &\notin \mathsf{Ref}(M_3)\\
(a;\{\epsilon,\ b\};\emptyset) &\in \mathsf{Ref}(M_6) & (a;\{\epsilon,\ b\};\emptyset) &\notin \mathsf{Ref}(M_5)
\end{aligned}$$

## 8 Conclusions and future work

In this paper we have studied the theory of safety and liveness for CCS extended with controlled reversibility, named $\text{CCS}_{\text{roll}}$. We developed characterisations of safety and liveness preorders which are based only on forward transitions (as in [6]), thus providing a simple proof technique for these preorders.

In $\text{CCS}_{\text{roll}}$ reductions are temporally ordered, forming a total order, and rollback always returns a system to a past state. In other reversibility calculi, reductions are causally ordered, and rollback is more involved [3, 16, 13]. We have opted for this form of reversibility for simplicity, in order to develop a theory of weak reduction barbed congruence for reversible languages. The adaptation of our theory to an uncontrolled reversibility is immediate because in such a setting all system states are reachable from each-other, thus may-testing implies should-testing (liveness and safety collapse to the same relation).

We have identified two key properties which, if true in any reversible language then our safety and liveness theory applies to that language. The first property can be shown to apply in a language with controlled, causal reversibility (e.g., [12]) and therefore our safety theory applies to such languages. The problem of determining whether the second property applies to [12] was insurmountable to us and remains an open question which we hope future work will be able to answer.

We have showed that with respect to safety, total reversibility is a conservative extension of CCS. With respect to liveness, however, adding total reversibility to CCS distinguishes more systems. The characterisations we have developed for reversibility are fundamentally different than those for communicating transactions [7], illuminating the difference between the two constructs. To our knowledge, this work provides the first characterisations of safety, liveness and testing for reversible calculi.

## A Omitted Definitions and Lemmas

**Definition 15 (Configuration free keys).** *The set of keys of a configuration $M$, written $\mathsf{keys}(M)$, is inductively defined as follows:*

$$\begin{aligned}
&\mathsf{keys}(\mathbb{0}) = \mathbb{0} && \mathsf{keys}(\nu a.M) = \mathsf{keys}(M)\\
&\mathsf{keys}(M \parallel N) = \mathsf{keys}(M) \cup \mathsf{keys}(N) && \mathsf{keys}(k{:}P) = \{k\} \cup \mathsf{keys}(P)\\
&\mathsf{keys}([k_0{:}P;\, k]) = \{k_0, k\} \cup \mathsf{keys}(P) && \mathsf{keys}(P \parallel Q) = \mathsf{keys}(P) \cup \mathsf{keys}(Q)\\
&\mathsf{keys}(\sum_{i\in I} \alpha_i(\gamma).P_i) = \bigcup_{i\in I} \mathsf{keys}(P_i) && \mathsf{keys}(\mathtt{rl}\langle\gamma\rangle) = \emptyset\\
&\mathsf{keys}(\mathtt{rl}\langle k\rangle) = \{k\} && \mathsf{keys}(\nu a.P) = \mathsf{keys}(P)\\
&\mathsf{keys}(\mathtt{rec}X(\gamma).P) = \mathsf{keys}(P) && \mathsf{keys}(X) = \emptyset
\end{aligned}$$

**Lemma 7.** *If $M \xhookrightarrow{\langle k\rangle} M'$ and $M \xhookrightarrow{\langle k\rangle} M''$ then $M' = M''$.*

*Proof.* By induction on $M$.

**Lemma 8.**

1. *If $M_1 \equiv N_1 \xhookrightarrow{\langle k\rangle} N_2$ then there exists $M_2$ such that $M_1 \xhookrightarrow{\langle k\rangle} M_2 \equiv N_2$.*
2. *If $M_1 \doteq N_1 \xhookrightarrow{\langle k\rangle} N_2$ then there exists $M_2$ such that $M_1 \xhookrightarrow{\langle k\rangle} M_2 \doteq N_2$.*

*Proof.* 1. By induction on the derivation of $M_1 \equiv N_1$, with case analysis on the last applied axiom. We consider some key cases.

$M_1 = N_1'$ **and** $N_1 = N_1' \parallel \mathbb{0}$ . By hypothesis we have that $N_1' \parallel \mathbb{0} \xhookrightarrow{\langle k\rangle} N_2'' \parallel \mathbb{0}$, which by rule RLPAR implies that $N_1' \xhookrightarrow{\langle k\rangle} N_2''$, and we can conclude by noticing that $N_2'' \equiv N_2'' \parallel \mathbb{0}$

$M_1 = M' \parallel N'$ **and** $N_1 = N' \parallel M'$ . By hypothesis we have that $N' \parallel M' \xhookrightarrow{\langle k\rangle} N_1' \parallel M_1'$, with $M' \xhookrightarrow{\langle k\rangle} M_1'$ and $N' \xhookrightarrow{\langle k\rangle} N_1'$. Hence also $M' \parallel N' \xhookrightarrow{\langle k\rangle} M_1' \parallel N_1'$ with $M_1' \parallel N_1' \equiv N_1' \parallel M_1'$, as desired.

$M_1 = k'{:}\nu a.P$ **and** $N_1 = \nu a.k'{:}P$ . By hypothesis, we have that $N_1 \xhookrightarrow{\langle k\rangle} N_2$. We have to distinguish two cases, either $k = k'$ or $k \neq k'$. In the first case we have that $\nu a.k'{:}P \xhookrightarrow{\langle k\rangle} \nu a.\mathbb{0}$ and $k'{:}\nu a.P \xhookrightarrow{\langle k\rangle} \mathbb{0}$ and we can conclude by noticing that $\nu a.\mathbb{0} \equiv \mathbb{0}$. The second case trivially holds.

$M_1 = k{:}(P \parallel Q)$ **and** $N_1 = k{:}P \parallel k{:}Q$ . By hypothesis, we have that $N_1 \xhookrightarrow{\langle k\rangle} N_2$. We have to distinguish two cases, either $k = k'$ or $k \neq k'$. In the first case we have that $k{:}P \parallel k{:}Q \xhookrightarrow{\langle k\rangle} \mathbb{0} \parallel \mathbb{0}$ since $k{:}P \xhookrightarrow{\langle k\rangle} \mathbb{0}$ and $k{:}Q \xhookrightarrow{\langle k\rangle} \mathbb{0}$. But also $k{:}(P \parallel Q) \xhookrightarrow{\langle k\rangle} \mathbb{0}$ and we can conclude by noticing that $\mathbb{0} \equiv \mathbb{0} \parallel \mathbb{0}$.

2. By induction on the length of the derivation of $M_1 \doteq N_1$. The proof is similar to the previous case. □

**Lemma 9.** *For any key $k$ and term $M$ such that $k \sharp M$, $M \xhookrightarrow{\langle k\rangle} M$.*

*Proof.* By structural induction on $M$.

**Lemma 10 (Forward-Backward).** *If $D \vdash M \xrightharpoonup{(k)} D' \vdash M'$ then there exists $M''$ such that $D' \vdash M' \xhookrightarrow{\langle k \rangle} D \vdash M''$ and $M \equiv M''$.*

*Proof.* By induction on $M \xrightharpoonup{(k)} M'$, using lemma 8 in the case of REQV and lemma 9 in the case of RPAR.

**Lemma 11.** *If $M \xrightharpoonup{(k)} M'$ then either*

$$\begin{array}{rlr} M & \equiv \nu\widetilde{a}.\,(k_0{:}P \parallel N) & P \xmapsto{\tau(k)} P' \\ M' & \equiv \nu\widetilde{a}.\,(k{:}P' \parallel [k_0{:}P;\ k] \parallel N) & k \sharp M, k_0, P, N \end{array}$$

*or*

$$\begin{array}{ll} M \equiv \nu\widetilde{a}.\,(k_1{:}P \parallel k_2{:}Q \parallel N) & P \xmapsto{a(k)} P' \\ M' \equiv \nu\widetilde{a}.\,(k{:}P' \parallel k{:}Q' \parallel [k_1{:}P;\ k] \parallel [k_2{:}Q;\ k] \parallel N) & Q \xmapsto{a(k)} Q' \quad k \sharp M, k_1, P, k_2, Q, N \end{array}$$

*for some $\widetilde{a}, \widetilde{k}, P, Q, P', Q', N$.*

*Proof.* By induction on the transition $M \xrightharpoonup{(k)} M'$.

**Lemma 12.** *If $M \xrightharpoonup{(k)} M'$ then $\mathsf{keys}(M') \subseteq \mathsf{keys}(M) \cup \{k\}$.*

*Proof.* By induction on the transition.

**Lemma 13.** *If $M \xhookrightarrow{\langle k \rangle} M'$ then $\mathsf{keys}(M') \subseteq \mathsf{keys}(M) \setminus \{k\}$.*

*Proof.* By induction on the transition.

**Lemma 14.** *If $M \equiv N$ then $\mathsf{keys}(M) = \mathsf{keys}(N)$.*

*Proof.* By induction on the derivation $M \equiv N$.

**Lemma 15.** *Let $M \xhookrightarrow{\langle k \rangle} M'$ and $N \in M'$ with $N = k{:}P$ or $[\mu;\ k]$, for some $k$, $\mu$, and $P$. Then it must be $N \in M$.*

*Proof.* By induction on the transition.

**Lemma 16.** *Let $M \xhookrightarrow{\langle k \rangle} M'$ and $M_1 \in M$ and $k \sharp M_1$. Then $M_1 \in M'$.*

*Proof.* By induction on the rollback transition.

**Lemma 17.** *If $\mathsf{wf}(D \vdash M)$ and $D \vdash M \xhookrightarrow{\langle k \rangle} D' \vdash M'$ then $D' \vdash M' \rightharpoonup^{*} D \vdash M$.*

*Proof.* By RLSYS from fig. 3 we have

$$M \xhookrightarrow{\langle k_1 \rangle} M_1 \cdots \xhookrightarrow{\langle k_n \rangle} M_n = M' \qquad D = (k_1 \prec \ldots \prec k_n \prec D')$$

By definition 2:

$$M \xhookrightarrow{\langle k_1 \rangle} M_1' \cdots \xhookrightarrow{\langle k_n \rangle} M_n'$$
$$M \xleftarrow{(k_1)} M_1' \cdots \xleftarrow{(k_n)} M_n'$$

By lemma 7, $M_1 = M_1', \ldots, M_n = M_n' = M'$, from which the result follows.

**Corollary 1.** *If* $\mathsf{wf}(D \vdash M)$ *and* $D \vdash M \hookrightarrow D' \vdash M'$ *then* $D' \vdash M' \rightharpoonup^* D \vdash M$.

Keys in well-formed systems are nested according to the causality relation.

**Lemma 18 (Syntactic Causality).** *If* $\mathsf{wf}(D \vdash M)$ *then*

1. Memory Causality: *If* $[\mu;\, l] \in M$ *and* $k \in \mathsf{keys}(\mu)$ *then* $D \vdash l \prec^+ k$.
2. Process Causality: *If* $l{:}P \in M$ *and* $k \in \mathsf{keys}(P)$ *then* $D \vdash l \prec^* k$.

Moreover, well-formed systems can roll back any memory, and the memories they contain are products of binary communication.

**Lemma 19 (Roll Back Any Memory).** *If* $\mathsf{wf}(D{\vdash}M)$ *and* $[\mu;\, k] \in M$ *then* $D{\vdash}M \xhookrightarrow{\langle k \rangle} D' \vdash M'$*, for some* $D'$ *and* $M'$.

**Lemma 20 (Binary Communication).** *If* $\mathsf{wf}(D{\vdash}M)$ *and* $M \equiv \nu\widetilde{a}.([\mu_1;\, k] \parallel [\mu_2;\, k] \parallel N)$ *then for all* $[\mu;\, l] \in N$, $k \neq l$.

**Lemma 21 (Preservation of Well-Formedness).** *Let* $\mathsf{wf}(D \vdash M)$*; then:*

1. *If* $M \equiv N$ *then* $\mathsf{wf}(D \vdash N)$.
2. *If* $M = \nu a.N$ *then* $\mathsf{wf}(D \vdash N)$.
3. *If* $D \vdash M \xrightharpoonup{(k)} D' \vdash M'$ *then* $\mathsf{wf}(D' \vdash M')$.
4. *if* $D \vdash M \xhookrightarrow{\langle k \rangle} D' \vdash M'$ *then* $\mathsf{wf}(D' \vdash M')$.
5. *If* $D \vdash M \to D' \vdash M'$ *then* $\mathsf{wf}(D' \vdash M')$.

*Proof.* The first property follows from lemmas 8 and 14. The second property follows by the definition of keys and rule RL$\nu$ of fig. 3.

To prove the third property we derive $D' = k \prec D$ by RSYS of fig. 2. We need to show that for some $M''$: $\mathsf{keys}(M') \subseteq \mathsf{keys}(D')$ and $M' \xhookrightarrow{\langle k \rangle} M''$ and $M'' \xrightharpoonup{(k)} M'$ and $\mathsf{wf}(D \vdash M'')$. The first follows from lemma 12. The next two follow from lemma 10 and rule REQV of fig. 2; the last follows from the first property of the lemma.

The fourth property follows by definition of well-formedness and rule BW of fig. 3, using an induction on the number of individual rollback transitions. The last property follows from the previous two.

**Lemma 22.** *Let* $D \vdash M \xrightharpoonup{(k)}\xhookrightarrow{\langle k \rangle} D' \vdash M'$. *Then* $D = D'$ *and* $M \equiv M'$.

*Proof.* By lemmas 1 and 10.

**Lemma 23.** *Let* $D \vdash M \xrightarrow{\langle k \rangle}\!\!\hookrightarrow^{\langle l \rangle} D' \vdash M'$ *and* $k \not\sharp\, l$. *Then there exists* $M'' \equiv M'$ *such that* $D \vdash M \xhookrightarrow{\langle l \rangle} D' \vdash M''$.

*Proof.* We have $D \vdash M \xrightharpoonup{\langle k \rangle} D_1 \vdash M_1 \xhookrightarrow{\langle l \rangle} D' \vdash M'$. By rule RSYS of fig. 2 we have $D_1 = k \prec D$. By lemma 10, there exists $M_2 \equiv M$ such that $D_1 \vdash M_1 \xhookrightarrow{\langle k \rangle} D \vdash M_2$.

By rule RLSYS of fig. 3 and $D_1 \vdash M_1 \xhookrightarrow{\langle l \rangle} D' \vdash M'$ we get a sequence of keys $\widetilde{k}$ such that $D_1 = \widetilde{k} \prec l \prec D'$ and $M_1 \xhookrightarrow{\langle \widetilde{k} \rangle}\xhookrightarrow{\langle l \rangle} M'$. Because $D_1 = k \prec D$ and $k \not\sharp\, l$, the sequence $\widetilde{k}$ must have at least one element, $k$. Thus there exists $M_2'$ and $\widetilde{k}'$ such that $D_1 = k \prec \widetilde{k}' \prec l \prec D'$ and $M_1 \xhookrightarrow{\langle k \rangle} M_2' \xhookrightarrow{\langle \widetilde{k}' \rangle}\xhookrightarrow{\langle l \rangle} M'$. From lemma 1, $M_2' = M_2 \equiv M$. By repeated applications of lemma 8 ($M_2' \equiv M$) we get $M \xhookrightarrow{\langle \widetilde{k}' \rangle}\xhookrightarrow{\langle l \rangle} M''$, for some $M'' \equiv M'$. The result follows from rule RLSYS.

**Corollary 2.** *Let* $D \vdash M \xrightharpoonup{\langle \widetilde{k} \rangle}\xhookrightarrow{\langle l \rangle} D' \vdash M'$ *and* $\widetilde{k} \not\sharp\, l$. *Then there exists* $M'' \equiv M'$ *such that* $D \vdash M \xhookrightarrow{\langle l \rangle} D' \vdash M''$.

**Lemma 24.** *Let* $D \vdash M \xrightharpoonup{\langle \widetilde{k} \rangle}\xhookrightarrow{\langle l \rangle} D' \vdash M'$. *Then one of the following holds:*

1. $D \vdash M \xhookrightarrow{\langle l \rangle} D' \vdash M'' \equiv M'$ *and* $l \in D$, *for some* $M''$, *or*
2. $D \vdash M \xrightharpoonup{\langle \widetilde{k}' \rangle} D' \vdash M'' \equiv M'$ *and* $\widetilde{k} = \widetilde{k}', l, \widetilde{k}''$, *for some* $M''$, $\widetilde{k}'$, $\widetilde{k}''$.

*Proof.* By cases on $\widetilde{k}$.

If $\widetilde{k} \not\sharp\, l$ then the first property follows from corollary 2.

Otherwise there exist $\widetilde{k}'$, $\widetilde{k}''$ such that $k = \widetilde{k}', l, \widetilde{k}''$ and $\widetilde{k}'' \not\sharp\, l$. In this case the result follows from corollary 2 and lemma 22.

**Lemma 25.** *Let* $D \vdash M \xhookrightarrow{\langle k \rangle}\xhookrightarrow{\langle l \rangle} D' \vdash M'$. *Then* $D \vdash M \xhookrightarrow{\langle l \rangle} D' \vdash M'$.

*Proof.* From rule RLSYS of fig. 3 we get $\widetilde{k}$, $\widetilde{l}$ such that $D = \widetilde{k} \prec k \prec \widetilde{l} \prec l \prec D'$ and $M \xhookrightarrow{\langle \widetilde{k} \rangle}\xhookrightarrow{\langle k \rangle}\xhookrightarrow{\langle \widetilde{l} \rangle}\xhookrightarrow{\langle l \rangle} M'$. The lemma is proven by an application of RLSYS.

**Property 3** *Let* $\mathsf{wf}(\widetilde{k} \prec D \vdash M_1)$ *and* $M_1 \xhookrightarrow{\langle \widetilde{k} \rangle} M_2$*; then* $D \vdash M_2 \rightharpoonup^* \widetilde{k} \prec D \vdash M_1$.

*Proof.* By induction on the number of $\widetilde{k}$. The base case, $\widetilde{k} = \emptyset$ banally holds. In the inductive case we have

$$k_n \prec \ldots \prec k_2 \prec k_1 \prec D \vdash M_1 \xhookrightarrow{\langle k_n, \ldots, k_2 \rangle} k_1 \prec D \vdash M_2 \xhookrightarrow{\langle k_1 \rangle} D \vdash M_1$$

By applying inductive hypothesis on the reduction $\xhookrightarrow{\langle k_n, \ldots k_2 \rangle}$, we obtain

$$k_n \prec \ldots \prec k_2 \prec k_1 \prec D \vdash M_1 \xhookrightarrow{\langle k_n, \ldots, k_2 \rangle} k_1 \prec D \vdash M_2 \rightharpoonup^* \widetilde{k} \prec D \vdash M$$

By Lemma 21 we have that $\mathsf{wf}(k_1 \prec D \vdash M_2)$, and by definition of $\mathsf{wf}(\cdot)$ we have that $k_1 \prec D \vdash M_2 \xhookrightarrow{\langle k_1 \rangle} D \vdash M_1 \rightharpoonup k_1 \prec D \vdash M_2$ and we can conclude by noticing that

$$k_n \prec \ldots \prec k_2 \prec k_1 \prec D \vdash M_1 \xhookrightarrow{\langle k_n, \ldots, k_2 \rangle} k_1 \prec D \vdash M_2 \xhookrightarrow{\langle k_1 \rangle} D \vdash M \rightharpoonup k_1 \prec D \vdash M_2 \rightharpoonup^* \widetilde{k} \prec D \vdash M$$

**Corollary 3.** *Let* $\mathsf{wf}(D_1 \vdash M_1)$ *and* $D_1 \vdash M_1 \xhookrightarrow{\langle k \rangle} D_2 \vdash M_2$*; then* $D_2 \vdash M_2 \rightharpoonup^* D_1 \vdash M_1$.

**Lemma 26.** *If* $D_1 \vdash M_1 \rightharpoonup^* D_2 \prec D_1 \vdash M_2 \xhookrightarrow{\langle l \rangle} D_3 \vdash M_3$ *and* $l \in D_2$ *then* $D_1 \vdash M_1 \rightharpoonup^* D_3 \vdash M_3$.

*Proof.* Let $D_2 = \widetilde{k_1} \prec l \prec \widetilde{k_2}$. We proceed by induction on the size of $\widetilde{k_2}$. In the base case, $\widetilde{k_2} = \emptyset$ we have that $D_1 \vdash M_1 \xrightharpoonup{(l)} \xrightharpoonup{(\widetilde{k_1})} \widetilde{k_1} \prec l \vdash M_2$. By Lemma 2 we have that there exists $M_1' \equiv M_1$ such that $\widetilde{k_1} \prec l \vdash M_2 \xhookrightarrow{\langle l \rangle} D_1 \vdash M_1'$. Moreover since rollback is deterministic, and since $\widetilde{k_1} \prec l \vdash M_2 \xhookrightarrow{\langle l \rangle} D_3 \vdash M_3$ by Lemma 1 we have that $D_1 = D_3$ and $M_1' = M_3$. We can conclude since $\rightharpoonup$ is closed under $\equiv$.

In the inductive case, let $\widetilde{k_2} = \widetilde{k_3} \prec k_2'$, we have that

$$D_1 \vdash M_1 \xrightharpoonup{(k_2')} k_2' \prec D_1 \vdash M' \xrightharpoonup{(\widetilde{k_3})} \widetilde{k_3} \prec k_2' \prec D_1 \vdash M'' \xrightharpoonup{(l)} \xrightharpoonup{(\widetilde{k_1})} D_2 \prec D_1 \vdash M_2 \xhookrightarrow{\langle l \rangle} D_3 \vdash M_3$$

By inductive hypothesis we have that $k_2' \prec D_1 \vdash M' \xrightharpoonup{(k_2')} k_2' \prec D_1 \vdash M' \rightharpoonup^* D_3 \vdash M_3$, and since $D_1 \vdash M_1 \xrightharpoonup{(k_2')} k_2' \prec D_1 \vdash M'$ we have that $D_1 \vdash M_1 \rightharpoonup^* D_3 \vdash M_3$, as desired.

**Lemma 27.** *Let* $D_1 \vdash M_1 \rightharpoonup^* D_2 \vdash M_2 \hookrightarrow D_3 \vdash M_3$. *One of the following is true:*

1. $D_3$ is smaller than $D_1$: *there exists non-empty* $D_4$ *such that* $D_1 = D_4 \prec D_3$*, or*
2. $D_3$ is larger than or equal to $D_1$: $D_1 \vdash M_1 \rightharpoonup^* D_3 \vdash M_3$.

*Proof.* By induction on the number $n$ of steps $D_1 \vdash M_1 \rightharpoonup^* D_2 \vdash M_2$. In the base case we have that $D_1 = D_2$ and $M_1 = M_2$. Moreover, since $D_2 \vdash M_2 \hookrightarrow D_3 \vdash M_3$, this implies that there exists a key $l \in D_2$ such that $D_2 = D_4' \prec l \prec D_3$ such that $D_2 \vdash M_2 \xhookrightarrow{\langle l \rangle} D_3 \vdash M_3$. And we have that 1 holds by setting $D_4 = D_4' \prec l$.

In the inductive case, we have that:

$$D_1 \vdash M_1 \xrightharpoonup{(k_1)} D_1' \vdash M_1' \xrightharpoonup{(\widetilde{k})} D_2 \vdash M_2 \hookrightarrow D_3 \vdash M_3$$

with $D_2 = \widetilde{k} \prec l \prec D_1$ and $D_1 = l_1 \prec D_1'$. By inductive hypothesis we have that if 1 holds, then $D_1' = D_4' \prec D_3$ with $D_4'$ non empty. And the property still holds by setting $D_4 = k_1 \prec D_4'$, as desired. By inductive hypothesis we have that if 2 holds, then $D_3$ is larger than or equal to $D_1'$: $D_1' \vdash M_1' \rightharpoonup^* D_3 \vdash M_3$. And the property still holds since $D_1 \vdash M_1 \xrightharpoonup{(k)} D_1' \vdash M_1' \rightharpoonup^* D_3 \vdash M_3$.

**Lemma 28.** *If* $D \vdash M \rightharpoonup^* D \vdash N$ *then* $N = M$.

Property 2 is a direct consequence of the following lemma.

**Lemma 29.** *Let* $\varepsilon \vdash M \rightharpoonup^* D \vdash N$ *and* $\mathsf{keys}(M) = \emptyset$*; there exist* $D_1$ *and* $N_1$*:*

1. $\varepsilon \vdash M \rightharpoonup^* D_1 \vdash N_1$
2. *if* $D_1 \vdash N_1 \rightharpoonup^* D' \vdash N'$ *then* $D \vdash N \rightarrow^* D' \vdash N'$
3. *if* $D \vdash N \rightarrow^* D' \vdash N'$ *then there exists* $N''$ *such that* $D_1 \vdash N_1 \rightharpoonup^* D' \vdash N'' \equiv N'$
4. *if* $D_1 \vdash N_1 \rightarrow^* D' \vdash N'$ *then there exists* $N''$ *such that* $D_1 \vdash N_1 \rightharpoonup^* D' \vdash N'' \equiv N'$

*Proof.* We expand the forward reduction in the assumption using rule Fw of fig. 2.

$$\varepsilon \vdash M \xrightharpoonup{(\widetilde{k})} D \vdash N \tag{4}$$

The following set contains the states reachable from $D \vdash N$ with arbitrary transitions:

$$\mathcal{S} = \{D' \vdash N' \quad | \quad D \vdash N \rightarrow^* D' \vdash N'\}$$

We take the earliest such state: let $(D_1 \vdash N_1) \in \mathcal{S}$ such that $D_1$ is a prefix of all states in $\mathcal{S}$. If there are multiple states with the same $D_1$ then we pick one of them.

We know that $\varepsilon \vdash M \rightharpoonup^* D \vdash N \rightarrow^* D_1 \vdash N_1$. Therefore, by property 1: $\varepsilon \vdash M \rightharpoonup^* D_1 \vdash N_1$, which proves the first conclusion of the lemma.

Because $D \vdash N \rightarrow^* D_1 \vdash N_1$, it follows that if $D_1 \vdash N_1 \rightharpoonup^* D_2 \vdash N_2$ then also $D \vdash N \rightarrow^* D_2 \vdash N_2$, which proves the second conclusion. For the same reason, the fourth conclusion follows from the third.

It remains to show the third conclusion.

We first establish that $D_1 \vdash N_1 \rightharpoonup^* D \vdash N' \equiv N$, for some $N'$. From lemma 3 one of the following holds:

1. $D \vdash N \rightharpoonup^* D_1 \vdash N_1' \equiv N_1$, for some $N_1'$.
   In this case, because $D_1$ is the smallest prefix of all $D$'s in $\mathcal{S}$, it must be that $D = D_1$ and the proof is completed by lemma 28 which gives us $N = N_1'$ and thus $D_1 \vdash N_1 \rightharpoonup^* D \vdash N_1 \equiv N_1' = N$.
2. $D \vdash N \xhookrightarrow{\langle l \rangle} D_0 \vdash N_0 \rightharpoonup^* D_1 \vdash N_1' \equiv N_1$, and $D \vdash N \rightarrow^* D_0 \vdash N_0' \equiv N_0$, for some $l \in D$, $D_0$, $N_1'$, $M_0$, $M_0'$.
   Because $(D_0 \vdash N_0') \in \mathcal{S}$ and we took $D_1$ to be the smallest prefix of all states in $\mathcal{S}$, it must be that $D_0 = D_1$. Thus, from lemma 28 $N_0 = N_1'$. From corollary 3, $D_1 \vdash N_1' \rightharpoonup^* D \vdash N$. Note that $\mathsf{wf}(\varepsilon \vdash M)$ and thus $\mathsf{wf}(D \vdash N)$ by lemma 21. Therefore, using rule Reqv, we derive that $D_1 \vdash N_1 \rightharpoonup^* D \vdash N' \equiv N$, for some $N'$.

Hence, $D_1 \vdash N_1 \rightharpoonup^* D \vdash N' \equiv N$, for some $N'$.

It now suffices to show that if $D_1 \vdash N_1 \rightarrow^* D' \vdash N'$ then $D_1 \vdash N_1 \rightharpoonup^* D' \vdash N'' \equiv N'$, for some $N''$. We prove this by induction on the number of reductions. The base case is trivial.

In the inductive case we have $D_1 \vdash N_1 \rightarrow^* D_2 \vdash N_2 \rightarrow D' \vdash N'$. By the induction hypothesis we get $D_1 \vdash N_1 \rightharpoonup^* D_2 \vdash N_2' \equiv N_2$, for some $N_2'$. Therefore, $D_2 \vdash N_2' \rightarrow D' \vdash N'' \equiv N'$, for some $N''$. If this $(\rightarrow)$ reduction is a forward transition then the proof is completed. Otherwise $D_1 \vdash N_1 \rightharpoonup^* D_2 \vdash N_2' \hookrightarrow D' \vdash N'' \equiv N'$. From lemma 27 we get that either:

$$\text{T}\alpha \quad \frac{P \xmapsto{\alpha(k)} P' \qquad k \sharp D}{D \models l{:}P \xrightarrow{\alpha(k)} (k \prec D) \models k{:}P' \parallel [l{:}P;\, k]}$$

$$\text{TPAR} \quad \frac{D \models M \xrightarrow{\alpha(k)} D' \models M' \qquad k \sharp N}{D \models M \parallel N \xrightarrow{\alpha(k)} D' \models M' \parallel N}$$

$$\text{TSYNC} \quad \frac{D \models M \xrightarrow{a(k)} D' \models M' \qquad D \models N \xrightarrow{\overline{a}(k)} D' \models N'}{D \models M \parallel N \xrightarrow{\tau(k)} D' \models M' \parallel N'}$$

$$\text{T}\nu \quad \frac{D \models M \xrightarrow{\alpha(k)} D' \models M' \qquad \alpha \sharp_n a}{D \models \nu a.M \xrightarrow{\alpha(k)} D' \models \nu a.M'}$$

$$\text{TEQV} \quad \frac{M \stackrel{\circ}{=} N \qquad D \models N \xrightarrow{\alpha(k)} D' \models N' \qquad N' \stackrel{\circ}{=} M'}{D \models M \xrightarrow{\alpha(k)} D' \models M'}$$

**Fig. 7.** LTS Transitions Based on Forward Moves

- there exists non-empty $D_3$ such that $D_1 = D_3 \prec D'$. In this case we have that $D'$ is a strict prefix of $D_1$. Moreover $(D' \vdash N'') \in \mathcal{S}$. However we took $D_1$ to be the smallest $D_1 \in \mathcal{S}$, which leads to contradiction. Thus this case is vacuously true.
- *$D_3$ is larger than or equal to $D_1$:* $D_1 \vdash M_1 \rightharpoonup^* D' \vdash N'' \equiv N'$, as needed. □

**Definition 16 (Well-Formed Configuration).** *$D \models M$ is well-formed ($\mathsf{wf}(D \models M)$) when*

1. $\mathsf{keys}(M) \subseteq \mathsf{keys}(D)$ *and*
2. *if $D = k \prec D'$ then there exist $\alpha$ and $M'$ such that $M \xhookrightarrow{\langle k \rangle} M'$ and $D' \models M' \xrightarrow{\alpha(k)} D \models M$ and $\mathsf{wf}(D' \models M')$.*

Well-formedness in the reduction semantics implies well-formedness in the LTS, but not the converse.

**Lemma 30.** *If $\mathsf{wf}(D \vdash M)$ then $\mathsf{wf}(D \models M)$.*

**Lemma 31.** *Let $\mathsf{wf}(D \models M)$ and $\mathsf{wf}(D \models N)$; then:*

1. $\mathsf{wf}(D \models \nu a.M)$ *and* $\mathsf{wf}(D \models M \parallel N)$.
2. *If $M = \nu a.M'$ then $\mathsf{wf}(D \models M')$.*
3. *If $M = M_1 \parallel M_2$ then $\mathsf{wf}(D \models M_1)$ and $\mathsf{wf}(D \models M_2)$.*
4. *If $M \equiv M'$ then $\mathsf{wf}(D \models M')$.*
5. *If $D \models M \xrightarrow{\alpha(k)} D' \vdash M'$ then $\mathsf{wf}(D' \vdash M')$.*

**Lemma 32 ($(\equiv)$ is a Strong Bisimulation).** *If $M \equiv N$ and $D \models M \xrightarrow{\alpha(k)} D' \models M'$ then there exists $N' \equiv M'$ such that $D \models N \xrightarrow{\alpha(k)} D' \models N'$.*

**Lemma 33.** *Let $D \models M \xrightarrow{\alpha(k)} D' \models M'$; then either*

1. $\alpha = \tau$ and

$$M \equiv \nu\widetilde{a}.\,(l{:}P \parallel N) \qquad M' \equiv \nu\widetilde{a}.\,([l{:}P;\, k] \parallel k{:}P' \parallel N)$$

with $P \xmapsto{\tau(k)} P'$ and $D' = k \prec D$ otherwise;
2. $\alpha = \tau$ and

$$M \equiv \nu\widetilde{a}.\,(l_1{:}P \parallel l_2{:}Q \parallel N) \qquad M' \equiv \nu\widetilde{a}.\,([l_1{:}P;\, k] \parallel [l_2{:}Q;\, k] \parallel k{:}P' \parallel Q')$$

with $P \xmapsto{a(k)} P'$, $Q \xmapsto{\overline{a}(k)} Q'$ and $D' = k \prec D$ otherwise;
3. $\alpha \neq \tau$ and

$$M \equiv \nu\widetilde{a}.\,(l{:}P \parallel N) \qquad M' \equiv \nu\widetilde{a}.\,([l{:}P;\, k] \parallel k{:}P' \parallel N)$$

with $P \xmapsto{\alpha(k)} P'$ and $D' = k \prec D$. □

**Corollary 4.** *$D \vdash M \rightharpoonup^* D' \vdash M'$ iff $D \models M \xrightarrow{\widetilde{\tau(k)}} D' \models M''$ with $M'' \equiv M'$.*

**Lemma 34.** *If $\mathcal{C} \xrightarrow{\alpha(k)} \mathcal{C}'$ then $k \sharp \mathcal{C}$.*

**Lemma 35.** *If $\mathcal{C} \xrightarrow{\alpha(k)} \mathcal{C}'$ and $p$ is a name permutation then $p\mathcal{C} \xrightarrow{\alpha(pk)} p\mathcal{C}'$.*

**Lemma 36 (Transition Weakening/Strengthening).** *Let $D \models M \xrightarrow{\alpha(k)} D' \models N$.*

1. *If $D = D_1 \prec l \prec D_2$ then $D_1 \prec D_2 \models M \xrightarrow{\alpha(k)} k \prec D_1 \prec D_2 \models N$.*
2. *If $D = D_1 \prec D_2$ and $l \sharp D, k$ then $D_1 \prec l \prec D_2 \models M \xrightarrow{\alpha(k)} k \prec D_1 \prec l \prec D_2 \models N$.*

**Lemma 37 (Canonical Traces are Typed).** *Let $t$ be a canonical trace with $t \sharp D$. There exists $D'$ such that $(D \vdash t \triangleright D')$.*

**Lemma 38 (Typed Traces are Canonical).** *If $(D \vdash t \triangleright D')$ then $t$ is canonical.*

**Lemma 39.** *If $(D \vdash t \triangleright D')$ then $t \sharp D$ and $\mathsf{keys}(D') = \mathsf{keys}(D) \cup \mathsf{keys}(t)$.*

Our characterisations of the safety and liveness preorders are based on *zipping* and un-zipping typed traces. The definition of zipping, shown in fig. 8, is straightforward. Moreover, trace typing is invariant to key permutation. LTS traces are canonical by construction and thus they are typable.

**Lemma 40.** *If $D_1 \parallel D_2 \twoheadrightarrow D$, then $\mathsf{keys}(D_1) \subseteq \mathsf{keys}(D)$ and $\mathsf{keys}(D_2) \subseteq \mathsf{keys}(D)$.*

**Lemma 41 (Zipping Inversion).** *If $D_1 \parallel D_2 \twoheadrightarrow (k \prec D)$ then one of the following holds:*

1. *$(k \prec D_1) \parallel D_2 \twoheadrightarrow k \prec D$,*
2. *$D_1 \parallel (k \prec D_2) \twoheadrightarrow k \prec D$, or*
3. *$(k \prec D_1) \parallel (k \prec D_2) \twoheadrightarrow k \prec D$.*

$$
\begin{array}{llr}
\varepsilon \parallel \varepsilon & \twoheadrightarrow \varepsilon & (\mathrm{Z}\varepsilon) \\
(k \prec D_1) \parallel D_2 & \twoheadrightarrow k \prec D \text{ if } D_1 \parallel D_2 \twoheadrightarrow D \text{ and } k \sharp D_1, D_2 & (\mathrm{ZL}) \\
D_1 \parallel (k \prec D_2) & \twoheadrightarrow k \prec D \text{ if } D_1 \parallel D_2 \twoheadrightarrow D \text{ and } k \sharp D_1, D_2 & (\mathrm{ZR}) \\
(k \prec D_1) \parallel (k \prec D_2) & \twoheadrightarrow k \prec D \text{ if } D_1 \parallel D_2 \twoheadrightarrow D \text{ and } k \sharp D_1, D_2 & (\mathrm{Z}\textsc{sync})
\end{array}
$$

$$
\mathrm{ZT}\epsilon\ \frac{D_1 \parallel D_2 \twoheadrightarrow D}{(D_1 \vdash \epsilon \triangleright D_1) \parallel (D_2 \vdash \epsilon \triangleright D_2) \twoheadrightarrow (D \vdash \epsilon \triangleright D)}
$$

$$
\mathrm{ZTL}\ \frac{(k \prec D_1 \vdash t_1 \triangleright D_1') \parallel (D_2 \vdash t_2 \triangleright D_2') \twoheadrightarrow (k \prec D \vdash t \triangleright D')}{(D_1 \vdash \alpha(k), t_1 \triangleright D_1') \parallel (D_2 \vdash t_2 \triangleright D_2') \twoheadrightarrow (D \vdash \alpha(k), t \triangleright D')}
$$

$$
\mathrm{ZTR}\ \frac{(D_1 \vdash t_1 \triangleright D_1') \parallel (k \prec D_2 \vdash t_2 \triangleright D_2') \twoheadrightarrow (k \prec D \vdash t \triangleright D')}{(D_1 \vdash t_1 \triangleright D_1') \parallel (D_2 \vdash \alpha(k), t_2 \triangleright D_2') \twoheadrightarrow (D \vdash \alpha(k), t \triangleright D')}
$$

$$
\mathrm{ZT}\textsc{Sync}\ \frac{(k \prec D_1 \vdash t_1 \triangleright D_1') \parallel (k \prec D_2 \vdash t_2 \triangleright D_2') \twoheadrightarrow (k \prec D \vdash t \triangleright D')}{(D_1 \vdash a(k), t_1 \triangleright D_1') \parallel (D_2 \vdash \overline{a}(k), t_2 \triangleright D_2') \twoheadrightarrow (D \vdash \tau(k), t \triangleright D')}
$$

**Fig. 8.** Zipping Traces.

**Lemma 42.** *If* $(D_1 \vdash t_1 \triangleright D_1') \parallel (D_2 \vdash t_2 \triangleright D_2') \twoheadrightarrow (D \vdash t \triangleright D_1')$ *then* $D_1 \parallel D_1 \twoheadrightarrow D$.

**Lemma 43 (Key Permutation).** *Let $p$ be a key permutation. Then*

1. *If* $(D \vdash t \triangleright D')$ *then* $(pD \vdash pt \triangleright pD')$.
2. *If* $\mathcal{C} \xrightarrow{t} \mathcal{C}'$ *then* $p\mathcal{C} \xrightarrow{pt} p\mathcal{C}'$.

**Lemma 44.** *Let* $D \models M \xrightarrow{t} D' \models M'$. *Then:*

1. $(D \vdash t \triangleright D')$.
2. *If* $(D \vdash t \triangleright D'')$ *then* $D' = D''$.

*Proof.* We prove the two properties by induction on the length $n$ of the trace $t$.

1. The base case $n = 0$ and $t = \epsilon$ trivially holds, since $D = D'$ and $(D \vdash \epsilon \triangleright D)$. In the inductive case, we have that $t = \alpha(k), t'$, with $D \models M \xrightarrow{\alpha(k)} D_1 \models M_1 \xrightarrow{t'} D' \models M'$. According to Definition 6, in order to show $(D \vdash \alpha(k), t' \triangleright D')$ we have to prove $(k \prec D \vdash t' \triangleright D')$. By inductive hypothesis on the derivation $D_1 \models M_1 \xrightarrow{t'} D' \models M'$ we have that $(D_1 \vdash t' \triangleright D')$ and since $D \models M \xrightarrow{\alpha(k)} D_1 \models M_1$ by Lemma 33 we have that $D_1 = k \prec D$ and we can conclude $(k \prec D \vdash t' \triangleright D')$.
2. The base case $n = 0$ and $t = \epsilon$ trivially holds. In the inductive case, we have $t = \alpha(k), t'$ and $D \models M \xrightarrow{\alpha(k)} D_1 \models M_1 \xrightarrow{t'} D'' \models M'$. By hypothesis we have that $(D \vdash \alpha(k), t' \triangleright D')$ and we can derive $(k \prec D \vdash t' \triangleright D')$. By Lemma 33 we have that $D \models M \xrightarrow{\alpha(k)} k \prec D \models M_1$, and since $k \prec D \models M_1 \xrightarrow{t'} D'' \models M'$, we can apply the inductive hypothesis and obtain that $D' = D''$, as desired. □

**Lemma 45 (Unzipping Transition).** *Let* $D \models M \parallel N \xrightarrow{\alpha(k)} D' \models R'$ *and* $D_1 \parallel D_2 \twoheadrightarrow D$ *and* $\mathsf{wf}(D_1 \models M)$ *and* $\mathsf{wf}(D_2 \models N)$. *Then one of the following is true*

1. $D_1 \models M \xrightarrow{\alpha(k)} D_1' \models M'$ *and* $D_1' \parallel D_2 \twoheadrightarrow D'$ *and* $R' = M' \parallel N$.
2. $D_2 \models N \xrightarrow{\alpha(k)} D_2' \models N'$ *and* $D_1 \parallel D_2' \twoheadrightarrow D'$ *and* $R' = M \parallel N'$.
3. $D_1 \models M \xrightarrow{a(k)} D_1' \models M'$ $D_2 \models N \xrightarrow{\overline{a}(k)} D_2' \models N'$ *and* $D_1' \parallel D_2' \twoheadrightarrow D'$ *and* $R' = M' \parallel N'$ *and* $\alpha = \tau$.

Two typed complementary traces can be zipped provided the keys annotating their $\tau$ actions do not overlap. Their zipped trace will contain only $\tau$ actions.

**Lemma 46.** *Let* $(D_1 \vdash t_1 \triangleright D_1')$ *and* $(D_2 \vdash t_2 \triangleright D_2')$ *and* $D_1 \parallel D_2 \twoheadrightarrow D_3$ *and* $\mathsf{obs}(t_1) = t_1'$ *and* $\mathsf{obs}(t_2) = t_2'$ *and* $\overline{t}_1' = t_2'$ *and* $\mathsf{keys}(t_1) \setminus \mathsf{keys}(t_1') \sharp D_2'$ *and* $\mathsf{keys}(t_2) \setminus \mathsf{keys}(t_2') \sharp D_1'$. *Then there exists* $t_3$, $D_3'$ *such that* $(D_1 \vdash t_1 \triangleright D_1') \parallel (D_2 \vdash t_2 \triangleright D_2') \twoheadrightarrow (D_3 \vdash t_3 \triangleright D_3')$ *and* $D_1' \parallel D_2' \twoheadrightarrow D_3'$ *and* $\mathsf{obs}(t_3) = \epsilon$.

The set $\mathsf{Tr}(\mathcal{C})$ is closed under the permutation of keys fresh from $\mathcal{C}$.

**Lemma 47.** *If* $t \in \mathsf{Tr}(\mathcal{C})$ *and* $p$ *is a permutation such that* $p \sharp \mathcal{C}$ *then* $pt \in \mathsf{Tr}(\mathcal{C})$.

**Theorem 5 (Soundness).** *For any initial systems* $M$ *and* $N$, $M \sqsubset\!\!\!\sim_{\mathsf{tr}} N$ *implies* $N \sqsubset\!\!\!\sim_{\mathsf{safe}} M$.

*Proof.* We assume $M \sqsubset\!\!\!\sim_{\mathsf{tr}} N$, for initial systems $M$ and $N$. According to definition 4, we need to show that for all tests $T$, $\varepsilon \vdash M \parallel \varepsilon{:}T \Downarrow_\omega$ implies $\varepsilon \vdash N \parallel \varepsilon{:}T \Downarrow_\omega$.

Let $\varepsilon \vdash M \parallel \varepsilon{:}T \Downarrow_\omega$, for test $T$. From definition 3, Let $\varepsilon \vdash M \parallel \varepsilon{:}T \to^* D \vdash O \downarrow_\omega$, for some system $O$. From property 1, the same system is reached with only forward reductions: $\varepsilon \vdash M \parallel \varepsilon{:}T \rightharpoonup^* D \vdash O \downarrow_\omega$. From corollary 4 we get $\varepsilon \models M \parallel \varepsilon{:}T \xrightarrow{t} D \models O' \equiv O$ for $t = \widetilde{k(\tau)}$.

Since $\varepsilon \parallel \varepsilon \twoheadrightarrow \varepsilon$ we unzip the trace $t$ by applying Proposition 1 and obtain:

$$\varepsilon \models M \xrightarrow{t_1} D_1 \models M_1 \qquad (\varepsilon \vdash t_1 \triangleright D_1) \parallel (\varepsilon \vdash t_2 \triangleright D_2) \twoheadrightarrow (\varepsilon \vdash t \triangleright D)$$
$$\varepsilon \models \varepsilon{:}T \xrightarrow{t_2} D_2 \models T_2 \qquad O' \mathrel{\mathring{=}} M_1 \parallel T_2 \qquad \mathsf{obs}(t_1) = \overline{\mathsf{obs}(t_2)}$$

Moreover, since $O' \downarrow_\omega$ and $\omega$ is only present in $T$, $T_2 \downarrow_\omega$. By definition 7, $\mathsf{obs}(t_1) \in \mathsf{Tr}(\varepsilon \models M)$, and by $M \sqsubset\!\!\!\sim_{\mathsf{tr}} N$, $\mathsf{obs}(t_1) \in \mathsf{Tr}(\varepsilon \models N)$. Therefore, there exists trace $t_3$ with $\mathsf{obs}(t_1) = \mathsf{obs}(t_3)$ such that $\varepsilon \models N \xrightarrow{t_3} D_3 \models N_3$ and, by lemma 44, $(\varepsilon \vdash t_3 \triangleright D_3)$. The keys in the observable part of $t_3$ are the same as in those of $t_1$. However, $t_3$ may contain $\tau$-actions whose keys overlap with the keys of $\tau$-actions in $t_2$. For this reason we invent a permutation $p$ which maps all $\tau$ keys of $t_3$ to fresh keys. Because $t_3 \sharp (\varepsilon \models N)$ (from lemma 39, $\mathsf{wf}(\varepsilon \models N)$) we get $p \sharp (\varepsilon \models N)$. Thus from lemma 43, $\varepsilon \models N \xrightarrow{t_4} D_4 \models N_4$, where $t_4 = pt_3$, $D_4 = pD_3$ and $N_4 = pN_3$. From lemma 44, $(\varepsilon \vdash t_4 \triangleright D_4)$. Moreover, because $p \sharp \mathsf{obs}(t_3)$, we have $\mathsf{obs}(t_1) = \mathsf{obs}(t_4) = \overline{\mathsf{obs}(t_2)}$, and by construction of $p$ and lemma 39: $\mathsf{keys}(t_4) \setminus \mathsf{keys}(\mathsf{obs}(t_4)) \sharp \mathsf{keys}(t_2) = \mathsf{keys}(D_2)$ and $\mathsf{keys}(t_2) \setminus \mathsf{keys}(\mathsf{obs}(t_2)) \sharp \mathsf{keys}(t_4) = \mathsf{keys}(D_4)$. Thus we can apply lemma 46 and obtain $(\varepsilon \vdash t_4 \triangleright D_4) \parallel (\varepsilon \vdash t_2 \triangleright D_2) \twoheadrightarrow (\varepsilon \vdash t' \triangleright D')$ with $\mathsf{obs}(t') = \epsilon$. By the Zipping proposition (Proposition 2) we get that for some $O''$, $\varepsilon \models N \parallel \varepsilon{:}T \xrightarrow{t'} D' \models O''$ with $O'' \mathrel{\mathring{=}} N_4 \parallel T_2$. Therefore, by corollary 4, $\varepsilon \vdash N \parallel \varepsilon{:}T \to^* D' \vdash O''' \equiv N_4 \parallel T_2 \downarrow_\omega$. Thus $\varepsilon \vdash N \parallel \varepsilon{:}T \Downarrow_\omega$.

**Theorem 6 (Completeness).** *For any initial systems $M$ and $N$, $M \sqsubseteq_{\mathsf{safe}} N$ implies $N \sqsubseteq_{\mathsf{tr}} M$.*

*Proof.* Let $M \sqsubseteq_{\mathsf{safe}} N$ for initial systems $M$ and $N$. We must show that $\mathsf{Tr}(N) \subseteq \mathsf{Tr}(M)$.

We take observable $t \in \mathsf{Tr}(N)$. By definition, $\varepsilon \models N \xrightarrow{t_1} D_1 \models N_1$, for some $D_1$, $T_1$, $t_1$ with $\mathsf{obs}(t_1) = t$. From lemma 4, $\varepsilon \models \mathsf{Test}^{\mathsf{s}}(t) \xrightarrow{\overline{t}} D_2 \models T_2{\downarrow_\omega}$, for some $D_2$, $T_2$. By lemma 44, $(\varepsilon \vdash t_1 \triangleright D_1)$ and $(\varepsilon \vdash \overline{t} \triangleright D_2)$. Moreover, $\mathsf{keys}(t_1) \setminus \mathsf{keys}(\mathsf{obs}(t_1)) \sharp \mathsf{keys}(\mathsf{obs}(t_1)) = \mathsf{keys}(t) = \mathsf{keys}(\overline{t})$ and by lemma 39, $\mathsf{keys}(\overline{t}) = \mathsf{keys}(D_2)$. Because $t$ is observable, $\mathsf{keys}(\overline{t}) \setminus \mathsf{keys}(\mathsf{obs}(\overline{t})) = \emptyset \sharp D_1$. Thus we can apply lemma 46 and obtain $(\varepsilon \vdash t_1 \triangleright D_1) \parallel (\varepsilon \vdash \overline{t} \triangleright D_2) \twoheadrightarrow (\varepsilon \vdash t_0 \triangleright D)$ for some $t_0, D$ with $\mathsf{obs}(t_0) = \epsilon$. By the Zipping proposition (Proposition 2) we get that for some $O$, $\varepsilon \models N \parallel \varepsilon{:}\mathsf{Test}^{\mathsf{s}}(t) \xrightarrow{t_0} D \models O$ with $O \mathrel{\dot{=}} N_1 \parallel T_2$. Therefore, by corollary 4, $\varepsilon \vdash N \parallel \varepsilon{:}\mathsf{Test}^{\mathsf{s}}(t) \rightharpoonup^* D' \vdash O' \equiv N_1 \parallel T_2{\downarrow_\omega}$, and $\varepsilon \vdash N \parallel \varepsilon{:}\mathsf{Test}^{\mathsf{s}}(t){\Downarrow_\omega}$.

By $M \sqsubseteq_{\mathsf{safe}} N$, we get $\varepsilon \vdash M \parallel \varepsilon{:}\mathsf{Test}^{\mathsf{s}}(t){\Downarrow_\omega}$, thus $\varepsilon \vdash M \parallel \varepsilon{:}\mathsf{Test}^{\mathsf{s}}(t) \rightarrow^* D' \vdash O''{\downarrow_\omega}$. From property 1, $\varepsilon \vdash M \parallel \varepsilon{:}\mathsf{Test}^{\mathsf{s}}(t) \rightharpoonup^* D' \vdash O''{\downarrow_\omega}$. Therefore, by corollary 4, $\varepsilon \vdash M \parallel \varepsilon{:}\mathsf{Test}^{\mathsf{s}}(t) \xrightarrow{t_0'} D' \vdash O''' \equiv O''{\downarrow_\omega}$, for some $t_0'$ and $O'''$ such that $\mathsf{obs}(t_0') = \epsilon$. By the Unzipping Proposition (Proposition 1) we obtain:

$$\begin{array}{lll} \varepsilon \models M \xrightarrow{t_3} D_3 \models M_3 & (\varepsilon \vdash t_3 \triangleright D_3) \parallel (\varepsilon \vdash t_4 \triangleright D_4) \twoheadrightarrow (\varepsilon \vdash t_0' \triangleright D') & \\ \varepsilon \models \varepsilon{:}\mathsf{Test}^{\mathsf{s}}(t) \xrightarrow{t_4} D_4 \models T_4 & O''' \mathrel{\dot{=}} M_4 \parallel T_4 & \mathsf{obs}(t_3) = \overline{\mathsf{obs}(t_4)} \end{array}$$

Therefore $\overline{\mathsf{obs}(t_4)} \in \mathsf{Tr}(M)$. From lemma 4, there exists permutation $p$ such that $pt = \overline{t}_4$. Because $t$ is observable, $\overline{\mathsf{obs}(t_4)} = pt$, and by lemma 47, $t \in \mathsf{Tr}(M)$ (note $ppt = t$).

**Lemma 48.** *Let $(t; V; W)$ be a refusal; then:*

1. *For all $k \sharp t$, there exists $D$ such that $\varepsilon \vdash \mathsf{Test}^{\mathsf{l}}(t; V; W) \xrightarrow{\overline{t}, \tau(k)} D \models \mathsf{Test}^{\mathsf{l}}(V; W)$.*
2. *For all $t \in V$ and $(k \prec D) \sharp t$, there exists $T$ such that $D \models \mathsf{Test}^{\mathsf{l}}(V; W) \xrightarrow{\tau(k), \overline{t}} D' \models T$.*
3. *If $D \models \mathsf{Test}^{\mathsf{l}}(V; W) \xrightarrow{\alpha(k), t} D' \models T$ then $\alpha = \tau$ and $\overline{t} \in V$ and $D' \models T{\downarrow_{\mathtt{rl}\langle k \rangle}}$.*
4. $\mathsf{Test}^{\mathsf{l}}(V; W) \not\downarrow_\omega$.
5. *For all $t \in W$ and $D \sharp t$, there exists $T$ such that $D \models \mathsf{Test}^{\mathsf{l}}(V; W) \xrightarrow{\overline{t}} D' \models T{\downarrow_\omega}$.*
6. *If $t \in W$ and $D \models \mathsf{Test}^{\mathsf{l}}(V; W) \xrightarrow{\overline{t}} D' \models T$ then $T{\downarrow_\omega}$.*
7. *If $D \models \mathsf{Test}^{\mathsf{l}}(V; W) \xrightarrow{t} D' \models T{\downarrow_\omega}$ then $\overline{t} \in W$.*

Refusal sets are closed under subset of their second and third elements.

**Lemma 49 (Subset and Key Permutation Closure of $\mathsf{Ref}$).** *Let $(t; V; W) \in \mathsf{Ref}(M)$ and $(t'; V'; W')$ is a refusal such that $V' \subseteq V$ and $W' \subseteq W$ and $t = pt'$ ($p$ a key permutation). Then $(t'; V'; W') \in \mathsf{Ref}(M)$.*

**Theorem 7 (Soundness).** *For initial systems, $M \sqsubseteq_{\mathsf{ref}} N$ implies $N \sqsubseteq_{\mathsf{live}} M$.*

*Proof.* Let $M \sqsubseteq_{\mathsf{ref}} N$ and $T$ be a liveness test. We need to show that $N$ shd $T$ implies $M$ shd $T$ (definition 5). We prove the contra-positive: $M \not\text{shd}\, T$ implies $N \not\text{shd}\, T$. Let $M \not\text{shd}\, T$. By definition 5, there exists $D_0 \vdash O_0$ such that

$$\varepsilon \vdash M \parallel \varepsilon{:}T \to^* D_0 \vdash O_0 \quad \text{and} \quad \forall (D' \vdash O').\ D_0 \vdash O_0 \to^* D' \vdash O' \ \text{implies}\ D' \vdash O' \not\downarrow_\omega$$

From property 1, the same system is reached with only forward reductions:

$$\varepsilon \vdash M \parallel \varepsilon{:}T \rightharpoonup^* D_0 \vdash O_0 \quad \text{and} \quad \forall (D' \vdash O').\ D_0 \vdash O_0 \to^* D' \vdash O' \ \text{implies}\ D' \vdash O' \not\downarrow_\omega$$

Note that from $D_0 \vdash O_0$ we can reach systems with forward or rollback transitions. However, we apply lemma 29 and get $D \vdash O$ such that

$$\begin{gathered} \varepsilon \vdash M \parallel \varepsilon{:}T \rightharpoonup^* D \vdash O \\ \forall (D' \vdash O').\ D \vdash O \to^* D' \vdash O' \ \text{implies}\ D' \vdash O' \not\downarrow_\omega \\ \forall (D' \vdash O').\ \exists O''.\ D \vdash O \to^* D' \vdash O' \ \text{implies}\ D \vdash O \rightharpoonup^* D' \vdash O'' \equiv O' \end{gathered} \quad (5)$$

Using corollary 4 we derive

$$\varepsilon \models M \parallel \varepsilon{:}T \xrightarrow{t} D \models O' \equiv O \qquad t = \widetilde{k(\tau)}$$

$$\forall t', (D' \vdash O').\ \mathsf{obs}(t') = \epsilon \ \text{and}\ D \models O \xrightarrow{t'} D' \vdash O' \ \text{implies}\ D' \models O' \not\downarrow_\omega \quad (6)$$

$$\forall t', (D' \vdash O').\ \mathsf{obs}(t') = \epsilon \ \text{and}\ D \models O \xrightarrow{t'} D' \vdash O' \ \text{implies}\ D' \models O' \not\downarrow_{\mathtt{rl}\langle D \rangle} \quad (7)$$

Note that (7) follows from (5). Since $\varepsilon \parallel \varepsilon \twoheadrightarrow \varepsilon$ we unzip the trace $t$ by applying Proposition 1 and obtain:

$$\varepsilon \models M \xrightarrow{t_1} D_1 \models M_1 \qquad (\varepsilon \vdash t_1 \triangleright D_1) \parallel (\varepsilon \vdash t_2 \triangleright D_2) \twoheadrightarrow (\varepsilon \vdash t \triangleright D) \quad (8)$$

$$\varepsilon \models \varepsilon{:}T \xrightarrow{t_2} D_2 \models T_2 \qquad O' \mathrel{\hat{=}} M_1 \parallel T_2 \qquad \mathsf{obs}(t_1) = \overline{\mathsf{obs}(t_2)} \quad (9)$$

We take

$$V = \{\mathsf{obs}(t) \mid \exists \mathcal{C}.\ D_2 \models T_2 \xrightarrow{\overline{t}} \mathcal{C}\} \quad W = \{\mathsf{obs}(t) \mid \exists \mathcal{C}.\ D_2 \models T_2 \xrightarrow{\overline{t}} \mathcal{C}\downarrow_\omega\} \cup \{\mathsf{obs}(t) \mid \exists \mathcal{C}.\ D_2 \models T_2 \xrightarrow{\overline{t}} \mathcal{C}\downarrow_{\mathtt{rl}\langle D_2 \rangle}\}$$

and argue that $(\mathsf{obs}(t_1); V; W) \in \mathsf{Ref}(M)$ according to definition 13. Because of (8) and (9), it suffices to show that $V \cap \mathsf{Roll}(D_1 \models M_1) = W \cap \mathsf{Tr}(D_1 \models M_1) = \emptyset$:

- $V \cap \mathsf{Roll}(D_1 \models M_1) = \emptyset$: By contradiction. Assume this is not the case. Then there exist $s_1$, $s_2$, $D_1'$, $D_2'$, $D'$, $O''$, $O'''$ such that $\mathsf{obs}(s_1) = \overline{\mathsf{obs}(s_2)} = s$ and $D_1 \models M_1 \xrightarrow{s_1} D_1' \models M_1'\downarrow_{\mathtt{rl}\langle D_1 \rangle}$ and $D_2 \models T_2 \xrightarrow{s_2} D_2' \models T_2'$ and $(D_1 \vdash s_1 \triangleright D_1') \parallel (D_2 \vdash s_2 \triangleright D_2') \twoheadrightarrow (D \vdash \widetilde{\tau(k)} \triangleright D')$ and $D \vdash M_1 \parallel T_2 \rightharpoonup^* D' \vdash O'' \xhookrightarrow{\langle l \rangle} D'' \vdash O'''$ with $l \in D$ (using lemmas 43 and 46 and Proposition 2). This contradicts (7). Thus it is necessary $V \cap \mathsf{Roll}(D_1 \models M_1) = \emptyset$.
- $W \cap \mathsf{Tr}(D_1 \models T_1) = \emptyset$: With a similar argument, this is necessary for (6) and (7) to hold.

By $M \sqsubseteq_{\mathsf{ref}} N$, we get $(\mathsf{obs}(t_1); V; W) \in \mathsf{Ref}(N)$. Using lemmas 39, 39, 43, 44 and 46 and Proposition 2 we derive that there exist $t_3$, $D_3$, $N_3$, $t'$, $D'$, $O''$ and $O'''$ such that $V \cap \mathsf{Roll}(D_3 \models N_3) = W \cap \mathsf{Tr}(D_3 \models N_3) = \emptyset$ and $\varepsilon \models N \xrightarrow{t_3} D_3 \models N_3$ and $\mathsf{obs}(t_3) = \mathsf{obs}(t_1)$ and $(\varepsilon \vdash t_3 \triangleright D_3) \parallel (\varepsilon \vdash t_2 \triangleright D_2) \twoheadrightarrow (\varepsilon \vdash t' \triangleright D')$ and $\mathsf{obs}(t') = \epsilon$ and $\varepsilon \models N \parallel \varepsilon{:}T \xrightarrow{t'} D' \models O'' \equiv N_3 \parallel T_2$ and $\varepsilon \vdash N \parallel \varepsilon{:}T \rightharpoonup^* D' \models O''' \equiv N_3 \parallel T_2$.

It remains to establish that $D' \vdash N_3 \parallel T_2 \not\Downarrow_\omega$.

We first show by contradiction that

$$\text{if } D' \vdash N_3 \parallel T_2 \to^* D'' \vdash O^{(4)} \text{ then } D' \vdash N_3 \parallel T_2 \rightharpoonup^* D'' \vdash O^{(4)} \qquad (10)$$

Suppose that $D' \vdash N_3 \parallel T_2 \to^* D' \vdash O^{(4)} \to D''' \vdash O^{(5)}$ is the smallest trace for which (10) does not hold. Then $D' \vdash N_3 \parallel T_2 \rightharpoonup^* D' \vdash O^{(4)} \xhookrightarrow{\langle l \rangle} D''' \vdash O^{(5)}$ and $l \in D'$. Thus $D' \vdash N_3 \parallel T_2 \rightharpoonup^* D' \vdash O^{(4)}\downarrow_{\mathtt{rl}\langle l \rangle}$.

By corollary 4, $D' \models N_3 \parallel T_2 \xrightarrow{\widetilde{\tau(k)}} D' \models O^{(6)}\downarrow_{\mathtt{rl}\langle l \rangle}$, for some $O^{(6)} \equiv O^{(4)}$. From Proposition 1 we obtain:

$$D_3 \models N_3 \xrightarrow{t_4} \mathcal{C}_4 \qquad D_2 \models T_2 \xrightarrow{t_5} \mathcal{C}_5 \qquad \mathsf{obs}(t_4) = \overline{\mathsf{obs}(t_5)}$$

and either $\mathcal{C}_4\downarrow_{\mathtt{rl}\langle D_3 \rangle}$ or $\mathcal{C}_5\downarrow_{\mathtt{rl}\langle D_2 \rangle}$. If it is the former, we observe that $\mathsf{obs}t_4 \in V$ by construction of $V$ and thus $V \cap \mathsf{Roll}(D_3 \models N_3) \neq \emptyset$, which is a contradiction. If it is the latter, we observe that in this case $\mathsf{obs}t_4 \in W$ by construction of $W$ and thus $Q \cap \mathsf{Tr}(D_3 \models N_3) \neq \emptyset$, which is again a contradiction. Therefore (10) must be true.

Due to (10) it remains to show that for all $D^{(7)}$ and $O^{(7)}$:

$$\text{if } D' \vdash N_3 \parallel T_2 \rightharpoonup^* D^{(7)} \vdash O^{(7)} \text{ then } D^{(7)} \vdash O^{(7)} \not\downarrow_\omega \qquad (11)$$

Again we prove this by contradiction. Assume $D' \vdash N_3 \parallel T_2 \rightharpoonup^* D^{(7)} \vdash O^{(7)}\downarrow_\omega$. By corollary 4, $D' \models N_3 \parallel T_2 \xrightarrow{\widetilde{\tau(k)}} D^{(7)} \models O^{(8)}\downarrow_\omega$, for some $O^{(8)} \equiv O^{(6)}$. From Proposition 1:

$$D_3 \models N_3 \xrightarrow{t_6} \mathcal{C}_6 \qquad D_2 \models T_2 \xrightarrow{t_7} \mathcal{C}_7 \qquad \mathsf{obs}(t_6) = \overline{\mathsf{obs}(t_7)}$$

and $\mathcal{C}_7\downarrow_\omega$. By construction of $W$, $\mathsf{obs}(t_6) \in W$. Therefore $W \cap \mathsf{Tr}D_3 \models N_3 \neq \emptyset$ which contradicts the definition of $(\mathsf{obs}(t_1); V; W) \in \mathsf{Ref}(N)$. Therefore it must be that $D' \vdash N_3 \parallel T_2 \not\Downarrow_\omega$ and thus $N \sqsubseteq_{\mathsf{live}} M$.

## B Omitted Proofs

*Proof (Proof of lemma 1 on p. 5).* By RLSYS of fig. 3 we get

$$D = (k_1 \prec \ldots \prec k_m \prec k \prec D') \qquad M \xhookrightarrow{\langle k_1 \rangle} M_1 \cdots \xhookrightarrow{\langle k_m \rangle} M_m \xhookrightarrow{\langle k \rangle} M'$$

because of $D \vdash M \xhookrightarrow{\langle k \rangle} D' \vdash M'$, and

$$D = (k'_1 \prec \ldots \prec k'_n \prec k \prec D'') \qquad M \xhookrightarrow{\langle k'_1 \rangle} M'_1 \cdots \xhookrightarrow{\langle k'_n \rangle} M'_n \xhookrightarrow{\langle k \rangle} M''$$

because of $D \vdash M \xhookrightarrow{\langle k \rangle} D'' \vdash M''$. By definition of dependency histories, $k$ appears only once in $D$, therefore $m = n$, $k_i = k'_i$ and $D' = D''$. We show $M' = M''$ by induction on $n$ and lemma 7.

*Proof (Proof of lemma 2 on p. 5).* By lemmas 8 and 10 and induction on $n$, we get $D' \vdash M' \xhookrightarrow{\langle l_n \rangle} \dots \xhookrightarrow{\langle l_1 \rangle}\xhookrightarrow{\langle k \rangle} D \vdash M'' \equiv M$. Thus $M' \xhookrightarrow{\langle l_n \rangle} \dots \xhookrightarrow{\langle l_1 \rangle}\xhookrightarrow{\langle k \rangle} M'' \equiv M$ and $D' = l_n \prec \dots \prec l_1 \prec k \prec D$, and we derive $D' \vdash M' \xhookrightarrow{\langle k \rangle} D \vdash M'' \equiv M$ as needed.

*Proof (Proof of lemma 3 on p. 7).* By induction on the number of rollback transitions in $D \vdash M \rightarrow^* D' \vdash M'$.

Base case: the reduction has no rollback transitions. The first property trivially holds.

Inductive case: the reduction can be decomposed (via the rules Fw and Bw of figs. 2 and 3, respectively) to:

$$D \vdash M \rightharpoonup^* \hookrightarrow D_1 \vdash M_1 \rightarrow^* D' \vdash M' \qquad \text{because} \qquad D \vdash M \xrightharpoonup{(\widetilde{k_1})}\xhookrightarrow{\langle l_1 \rangle} D_1 \vdash M_1 \tag{12}$$

The reduction $D_1 \vdash M_1 \rightarrow^* D' \vdash M'$ is strictly smaller than the original one, thus we can apply the induction hypothesis and get that one of the following holds:

1. $D_1 \vdash M_1 \rightharpoonup^* D' \vdash M'' \equiv M'$
   From lemma 24 and (12) we have the sub-cases:
   (a) $D \vdash M \xhookrightarrow{\langle l_1 \rangle} D_1 \vdash M'_1 \equiv M_1$, for some $l \in D$. From rule REQV of fig. 2, $D_1 \vdash M'_1 \rightharpoonup^* D' \vdash M''' \equiv M'' \equiv M'$, for some $M'''$. The proof is completed in this case by (12): $D \vdash M \rightarrow^* D_1 \vdash M_1 \equiv M'_1$.
   (b) $D \vdash M \xrightharpoonup{(\widetilde{k'_1})} D_1 \vdash M'_1 \equiv M_1$. The first property follows using again rule REQV.
2. $D_1 \vdash M_1 \xhookrightarrow{\langle l_0 \rangle} D_0 \vdash M_0 \rightharpoonup^* D' \vdash M'' \equiv M'$ and $D_1 \vdash M_1 \rightarrow^* D_0 \vdash M'_0 \equiv M_0$, for some $l_0 \in D_1$, $D_0$, $M''$, $M_0$, $M'_0$.
   By lemma 22, $D \vdash M \xrightharpoonup{(\widetilde{k_1})}\xhookrightarrow{\langle l_0 \rangle} D_0 \vdash M_0$.
   From lemma 24 we have the sub-cases:
   (a) $D \vdash M \xhookrightarrow{\langle l_0 \rangle} D_0 \vdash M''_0 \equiv M_0$ and $l_0 \in D$.
   From rule REQV of fig. 2, $D_0 \vdash M''_0 \rightharpoonup^* D' \vdash M''' \equiv M'' \equiv M'$, for some $M'''$. Moreover from (12), $D \vdash M \rightarrow^* D_1 \vdash M_1 \rightarrow^* D_0 \vdash M'_0 \equiv M_0 \equiv M''_0$, establishing the second conclusion of the lemma.
   (b) $D \vdash M \xrightharpoonup{(\widetilde{k'})} D_0 \vdash M''_0 \equiv M_0$.
   From rule REQV of fig. 2, $D_0 \vdash M''_0 \rightharpoonup^* D' \vdash M''' \equiv M'' \equiv M'$, for some $M'''$, establishing the first conclusion of the lemma. □

*Proof (Proof of Theorem 2 on p. 8).* In the $\Rightarrow$ direction, from hypothesis and rules Fw and RSYS of fig. 2, we get $M \xrightharpoonup{(k)} M'$. The result follows by induction on this derivation. The case REQV follows by the induction hypothesis and lemma 32.

The reverse direction follows from lemma 33.

*Proof (Proof of Proposition 1 on p. 9).* By induction on the length of the trace $t$ and then using the preceding lemma to take cases on the transition. The base case, $t = \epsilon$ banally holds. In the inductive case we have that $t = \alpha(k), s$ with:

$$D \models M \parallel N \xrightarrow{\tau(k)} D_s \models M_s \parallel N_s \xrightarrow{s} D' \models M' \parallel N'$$

We have three cases depending on who generated the action $\tau(k)$, either $M$, or $N$ or both.

In the first case we have that $D \models M \parallel N \xrightarrow{\tau(k)} D_s \models R_s$, with $R_s = M_s \parallel N_s$, $D \models M \xrightarrow{\tau(k)} D_s \models M_s$ and $N_s = N'$. By Lemma 45 we have that $D_1 \models M \xrightarrow{\tau(k)} D'_1 \models M_s$ and that $D'_1 \parallel D_2 \twoheadrightarrow D_s$. Since $\mathsf{obs}(t) = \epsilon$ and $t = \tau(k), s$ we have that $\mathsf{obs}(s) = \epsilon$. By inductive hypothesis on $D_s \models M_s \parallel N_n \xrightarrow{s} D' \models R'$ and $D'_1 \parallel D_2 \twoheadrightarrow D_s$ we have that there exist $M'$, $N'$, $D''_1$, $D''_2$, $s_1$ and $s_2$ such that

$$D_s \models M \xrightarrow{s_1} D''_1 \models M' \qquad (D'_1 \vdash s_1 \triangleright D''_1) \parallel (D_2 \vdash s_2 \triangleright D''_2) \twoheadrightarrow (D_s \vdash s \triangleright D')$$
$$D_s \models N \xrightarrow{s_2} D''_2 \models N' \qquad R' \mathrel{\mathring{=}} M' \parallel N' \qquad \mathsf{obs}(s_1) = \overline{\mathsf{obs}(t_2)}$$

by Lemma 33 we have that $D_s = k \prec D$ and $D'_1 = k \prec D_1$. We can conclude by noticing that:

–

$$D \models M \xrightarrow{\tau(k)} D_s \models M \xrightarrow{s_1} D''_1 \models M' \quad \text{implies} \quad D \models M \xrightarrow{\tau(k), s_1} D''_1 \models M'$$
$$D \models N \xrightarrow{\epsilon} D_s \models N \xrightarrow{s_2} D''_2 \models N' \quad \text{implies} \quad D \models N \xrightarrow{s_2} D''_2 \models N'$$

– By applying rule ZTL of Figure 8 we have that

$$(k \prec D_1 \vdash s_1 \triangleright D''_1) \parallel (D_2 \vdash s_2 \triangleright D''_2) \twoheadrightarrow (k \prec D \vdash s \triangleright D') \quad \text{implies}$$
$$(D_1 \vdash \alpha(k), s_1 \triangleright D''_1) \parallel (D_2 \vdash s_2 \triangleright D''_2) \twoheadrightarrow (D \vdash \alpha(k), s \triangleright D')$$

The other two cases are similar.

*Proof (Proof of Proposition 2 on p. 9).* By induction on $(D_1 \vdash t_1 \triangleright D'_1) \parallel (D_2 \vdash t_2 \triangleright D'_2) \twoheadrightarrow (D \vdash t \triangleright D')$. We have four cases, corresponding to the rules ZT$\epsilon$, ZTL, ZTR and ZSYNC of Figure 8. We will consider just the first two cases, the remaining ones are similar to the second case.

**ZT$\epsilon$** In this case we have that $t_1 = t_2 = t = \epsilon$, $D_1 = D'_1$ and $D_2 = D'_2$. Then the proposition banally holds.

**ZTL** In this case we have that $t_1 = \alpha(k), s_1$ :

$$D_1 \models M \xrightarrow{\alpha(k)} D_s \models M_s \xrightarrow{s_1} D'_1 \models M' \qquad D_2 \models N \xrightarrow{t_2} D'_2 \models N'$$

By Lemma 40 we have that $D_s = k \prec D_1$. By initial hypothesis we have that

$$(D_1 \vdash \alpha(k), s_1 \triangleright D'_1) \parallel (D_2 \vdash t_2 \triangleright D'_2) \twoheadrightarrow (D \vdash \alpha(k), s \triangleright D')$$

and by applying rule ZTL of Figure 8 we can derive

$$(k \prec D_1 \vdash s_1 \triangleright D_1') \parallel (D_2 \vdash t_2 \triangleright D_2') \twoheadrightarrow (k \prec D \vdash s \triangleright D')$$

We can now apply the inductive hypothesis and derive that $k \prec D \models M_s \parallel N \xrightarrow{s} D' \models R'$. We have now to show that $D \models M \parallel N \xrightarrow{\alpha(k)} k \prec D \models M_s \parallel N$. By applying Lemma 42 on the derivation $(k \prec D_1 \vdash s_1 \triangleright D_1') \parallel (D_2 \vdash t_2 \triangleright D_2') \twoheadrightarrow (k \prec D \vdash s \triangleright D')$, we have that $k \prec D_1 \parallel D_2 \twoheadrightarrow k \prec D$, and by Lemma 41 this implies that $D_1 \parallel D_2 \twoheadrightarrow D$. Since we are considering well formed system, $D_1 \models M \xrightarrow{\alpha(k)} D_s \models M_s$ implies that $\mathsf{keys}(M) \subseteq \mathsf{keys}(D_1)$ and by Lemma 40 we have that $\mathsf{keys}(D_1) \subseteq \mathsf{keys}(D)$ and hence $\mathsf{keys}(M) \subseteq \mathsf{keys}(D)$. We can now apply Lemma 36 and obtain $D \models M \xrightarrow{\alpha(k)} k \prec D \models M_s$ and by applying rule TPAR of Figure 5 we can derive $D \models M \parallel N \xrightarrow{\alpha(k)} k \prec D \models M_s \parallel N$. We can conclude by noticing that:

$$D \models M \parallel N \xrightarrow{\alpha(k)} k \prec D \models M_s \parallel N \xrightarrow{s} D' \models R'$$

□

*Proof (Proof of lemma 18 on p. 15).* In both cases, if $D \vdash k \prec^+ l$ then $D \vdash M$ would not be able to roll back $k$. Moreover, in the first case, if $k = l$ and we rolled back $k$ we would not be able to derive $D \vdash M$ with forward moves. Thus, by contradiction, both properties must hold.

*Proof (Proof of lemma 19 on p. 15).* By Condition 1 of definition 2, $\mathsf{wf}(D \vdash M)$ gives us $D = k_1 \prec \ldots \prec k_n \prec k \prec D'$, for some $k_1, \ldots, k_n$, and $D'$. By Condition 2 of the same definition we get $M \xhookrightarrow{\langle k_1 \rangle} M_1 \ldots \xhookrightarrow{\langle k_n \rangle} M_n \xhookrightarrow{\langle k \rangle} M'$, for some $M_1, \ldots M_n$, and $M'$. Therefore $D \vdash M \xhookrightarrow{\langle k \rangle} D' \vdash M'$ by Rule RLSYS of fig. 3.

*Proof (Proof of lemma 20 on p. 15).* By contradiction. Assume $[\mu;\, k] \in N$. By well-formedness (definition 2), $D = k_1 \prec \ldots \prec k_n \prec k \prec D'$ and $M \xhookrightarrow{\langle k_1 \rangle} M_1 \ldots \xhookrightarrow{\langle k_n \rangle} M_n \xhookrightarrow{\langle k \rangle} M'$ and $\mathsf{wf}(M_n)$. By lemma 18, $\widetilde{k}, k \sharp \mu_1, \mu_2, \mu$. Therefore, by lemma 16, $[\mu_1;\, k] \in M_n$, $[\mu_2;\, k] \in M_n$, and $[\mu;\, k] \in M_n$. Moreover, by lemma 13, $k \sharp M'$. Then it is not possible to derive $M' \xrightharpoonup{\langle k \rangle} M_n$, because forward reductions produce at most two memories (lemma 11). This means that $M_n$ is not well-produced, which is a contradiction. Thus it must be that for all $[\mu;\, l] \in N$, $k \neq l$.

*Proof (Proof of lemma 32 on p. 19).* By induction on the derivation $M \equiv N$, proving the lemma first for non-$\tau$ transitions. In the proof for $\tau$-transitions we use the lemma for non-$\tau$ transitions.

*Proof (Proof of lemma 45 on p. 21).* Since $D \models M \parallel N \xrightarrow{\alpha(k)} D' \models R'$, by Lemma 33 we have that $D' = k \prec D$. We then proceed by case analysis on the reduction $D \models M \parallel N \xrightarrow{\alpha(k)} D' \models R'$. We have three cases: $M$ by itself did the action, $N$ by itself did the action or both $M$ and $N$ contributed to the action.

1. We have that $D \models M \xrightarrow{\alpha(k)} D' \models M'$ implies $D \models M \parallel N \xrightarrow{\alpha(k)} D' \models M' \parallel N$, with $k \sharp D, M, N$. Since $D_1 \parallel D_2 \twoheadrightarrow D$ by Lemma 40 we have that $\mathsf{keys}(D_i) \subseteq \mathsf{keys}(D)$ with $i \in \{1, 2\}$, and since $k \sharp D$ then $k \sharp D_1, D_2$. We have then $D_1 \models M \xrightarrow{\alpha(k)} D_1' \models M'$ and by Lemma 33 we have that $D_1' = k \prec D_1$. By rule ZTL of fig. 8 we have that $k \prec D_1 \parallel D_2 \twoheadrightarrow k \prec D$, and we can conclude by noticing that $D' = k \prec D$.
2. this case is similar to the previous one.
3. We have that $D \models M \xrightarrow{a(k)} D' \models M'$ and $D \models N \xrightarrow{\overline{a}(k)} D' \models N'$ imply $D \models M \parallel N \xrightarrow{\tau(k)} D' \models M' \parallel N'$, with $k \sharp D, M, N$. Since $D_1 \parallel D_2 \twoheadrightarrow D$ by Lemma 40 we have that $\mathsf{keys}(D_i) \subseteq \mathsf{keys}(D)$ with $i \in \{1, 2\}$, and since $k \sharp D$ then $k \sharp D_1, D_2$. We have then $D_1 \models M \xrightarrow{a(k)} D_1' \models M'$ and $D_2 \models M \xrightarrow{a(k)} D_2' \models M'$ by Lemma 33 we have that $D_1' = k \prec D_1$ and $D_2' = k \prec D_2$. By rule ZTSYNC of fig. 8 we have that $k \prec D_1 \parallel k \prec D_2 \twoheadrightarrow k \prec D$, and we can conclude by noticing that $D' = k \prec D$. □

*Proof (Proof of lemma 46 on p. 22).* By induction on the structure of $t_1$ and $t_2$. In the base case the that $t_1 = t_2 = \epsilon$ and by definition of trace typing (Definition 6) we have that $D_1' = D_1$ and $D_2' = D_2$. By applying rule ZT$\epsilon$ we have that

$$(D_1 \vdash \epsilon \triangleright D_1) \parallel (D_2 \vdash \epsilon \triangleright D_2) \twoheadrightarrow (D_3 \vdash \epsilon \triangleright D_3')$$

and the lemma holds by setting $t_3 = \epsilon$ and $D_3' = D_3$. In the inductive hypothesis we have to do a case analysis on the form of $t_1$.

$t_1 = a(k), t_1'$ . Since $\mathsf{obs}(t_1) = t_1'$ and $\mathsf{obs}(t_2) = t_2'$ and $\overline{t}_1' = t_2'$, then there exists $\overline{a}(k)$ such that $t_2 = \widetilde{\tau(l)}, \overline{a}(k), t_2'$ and $\widetilde{l} \sharp D_1'$ which implies $\widetilde{l} \sharp D_1$. We have then that $D_1 \parallel \widetilde{l} \prec D_2 \twoheadrightarrow \widetilde{l} \prec D_3$. Since we have that $(\widetilde{l} \prec D_2 \vdash \overline{a}(k), t_2' \triangleright D)_2'$ by applying inductive hypothesis (condition on names holds) we obtain that

$$(D_1 \vdash a(k), t_1' \triangleright D_1') \parallel (\widetilde{l} \prec D_2 \vdash \overline{a}(k), t_2' \triangleright D_2') \twoheadrightarrow (\widetilde{l} \prec D_3 \vdash t_3' \triangleright D_3')$$

with $\mathsf{obs}(t_3') = \epsilon$ and $D_1' \parallel D_2' \twoheadrightarrow D_3$. Now, by applying rule ZTR $n$ times, with $n$ being the size of $\widetilde{\tau(l)}$ we obtain that

$$(D_1 \vdash a(k), t_1' \triangleright D_1') \parallel (D_2 \vdash \widetilde{\tau(l)}, \overline{a}(k), t_2' \triangleright D_2') \twoheadrightarrow (D_3 \vdash \widetilde{\tau(l)}, t_3' \triangleright D_3')$$

with since $D_1' \parallel D_2' \twoheadrightarrow D_3$. Since $\mathsf{obs}(\widetilde{\tau(l)}, t_3') = \mathsf{obs}(t_3') = \epsilon$ we can conclude.

$t_1 = \tau(k), t_1'$ . By initial hypothesis we have that $k \sharp D_2'$ and this implies $k \sharp D_2$. We have then that $k \prec D_1 \parallel D_2 \twoheadrightarrow k \prec D_3$. Since $(k \prec D_1 \vdash t_1' \triangleright D_1')$, by inductive hypothesis we have that

$$(k \prec D_1 \vdash t_1' \triangleright D_1') \parallel (D_2 \vdash t_2' \triangleright D_2') \twoheadrightarrow (k \prec D_3 \vdash t_3' \triangleright D_3')$$

with $D_1' \parallel D_2' \twoheadrightarrow D_3'$ and $\mathsf{obs}(t_3') = \epsilon$. We can now apply rule ZTR and obtain that:

$$(D_1 \vdash \tau(k), t_1' \triangleright D_1') \parallel (D_2 \vdash t_2' \triangleright D_2') \twoheadrightarrow (\prec D_3 \vdash \tau(k), t_3' \triangleright D_3')$$

and we can conclude by letting $t_3 = \tau(k), t_3'$ and by noticing that $\mathsf{obs}(t_3) = \mathsf{obs}(t_3') = \epsilon$, as desired.

$t_1 = \epsilon$ **and** $t_2 \neq \epsilon$ . By initial hypothesis, $t_2 = \tau(k), t'_2$ with $\mathsf{obs}(t'_2) = \epsilon$. Moreover, since $t_1 = \epsilon$ we have that $D_1 = D'_1$. Since $D_1 \parallel D_2 \twoheadrightarrow D_3$ and $k \sharp D'_1$ this implies also $k \sharp D_1$. We then have that $D_1 \parallel k \prec D_2 \twoheadrightarrow k \prec D_3$. We can then apply inductive hypothesis and obtain that

$$(D_1 \vdash \epsilon \rhd D_1) \parallel (k \prec D_2 \vdash t'_2 \rhd D'_2) \twoheadrightarrow (k \prec D_3 \vdash t'_3 \rhd D'_3)$$

with $D_1 \parallel D'_2 \twoheadrightarrow D'_3$ and $\mathsf{obs}(t_3) = \epsilon$. We can then apply rule ZTR and obtain

$$(D_1 \vdash \epsilon \rhd D_1) \parallel (D_2 \vdash \tau(k), t'_2 \rhd D'_2) \twoheadrightarrow (D_3 \vdash \tau(k), t'_3 \rhd D'_3)$$

as desired.

$t_1 = a(k), t'_1$ **and** $t_2 = \overline{a}(k), t'_2$ . Similar to the previous case, with the use of rule ZTSYNC to conclude.

*Proof (Proof of lemma 47 on p. 22).* Let $t \in \mathsf{Tr}(\mathcal{C})$ and $p$ be a permutation such that $p \sharp \mathcal{C}$. We have $\mathcal{C} \xrightarrow{t'} \mathcal{C}'$, for some $t'$ and $\mathcal{C}'$ with $\mathsf{obs}(t') = t$. From lemma 43, $p\mathcal{C} \xrightarrow{pt'} p\mathcal{C}'$. Moreover, $\mathsf{obs}(pt') = pt$ and $P\mathcal{C} = \mathcal{C}$ because $p \sharp \mathcal{C}$. Thus $pt \in \mathsf{Tr}(\mathcal{C})$.